\documentclass[onecolumn, preprintnumbers, showpacs, nofootinbib, aps]{revtex4-1}
\usepackage{graphicx}

\newcommand{\bi}{\bibitem}
\newcommand{\be}{\begin{eqnarray}}
\newcommand{\ee}{\end{eqnarray}}

\begin{document}

\title{Outflows from accreting super-spinars}

\author{Cosimo Bambi$^{\rm 1}$}
\email{cosimo.bambi@ipmu.jp}

\author{Tomohiro Harada$^{\rm 2}$}
\email{harada@rikkyo.ac.jp}

\author{Rohta Takahashi$^{\rm 3}$}
\email{rohta@riken.jp}

\author{Naoki Yoshida$^{\rm 1}$}
\email{naoki.yoshida@ipmu.jp}

\affiliation{
$^{\rm 1}$Institute for the Physics and Mathematics of the Universe, 
The University of Tokyo, Kashiwa, Chiba 277-8583, Japan\\
$^{\rm 2}$Department of Physics, Rikkyo University, 
Toshima, Tokyo 171-8501, Japan\\
$^{\rm 3}$Cosmic Radiation Laboratory, The Institute of Physical and 
Chemical Research, Wako, Saitama 351-0198, Japan}

\date{\today}

\preprint{IPMU10-0019}

\begin{abstract}
In this paper we continue our study on the accretion process onto
super-spinning Kerr objects with no event horizon (super-spinars). 
We discuss the counterpart of the Bondi accretion onto black holes. 
We first report the results of our
numerical simulations. We found a quasi steady-state configuration 
for any choice of the parameters of our model. The most interesting 
feature is the presence of hot outflows. Unlike jets and outflows 
produced around black holes, which are thought to be powered by 
magnetic fields and emitted from the poles, 
here the outflows are produced by the repulsive 
gravitational force at a small distance from the super-spinar 
and are ejected around the equatorial plane. In some circumstances, 
the amount of matter in the outflow is considerable,
which can indeed significantly reduce the gas mass accretion rate. 
Finally, we discuss a possible 
scenario of the accretion process in more realistic situations, 
which cannot be simulated by our code. 
\end{abstract}


\maketitle


\section{Introduction}

The actual nature of the final product of the gravitational collapse of 
matter is an outstanding and longstanding problem in general relativity
(GR). Under apparently reasonable assumptions, collapsing matter leads 
to the formation of space-time singularities~\cite{hawking}. Here there 
are two possibilities: $i)$ the singularity is hidden behind an event 
horizon and the final product is a black hole (BH), or $ii)$ the 
singularity is naked. Since space-times with naked singularities present 
pathologies, some form of the Cosmic Censorship Conjecture is usually 
assumed and naked singularities are forbidden~\cite{penrose}. Neglecting 
the electric charge, it turns out that, in four dimensions, the only 
stationary and asymptotically flat solution of the vacuum Einstein 
equation with an event horizon is the Kerr BH~\cite{carter, robinson}.

The Kerr metric is completely characterized by two parameters; that is, 
the mass $M$ and the spin $J$. The latter is often replaced by the Kerr 
parameter $a$, defined as $a = J/M$, or by the dimensionless Kerr parameter 
$a_* = a/M$. Using Boyer-Lindquist coordinates, the position of the 
horizon of a Kerr BH is given by
\be
r_H = M \left(1 + \sqrt{1 - a^2_*}\right) \, ,
\ee
which demands the well known constraint $|a_*| \le 1$. 
For $|a_*| > 1$, there is no horizon and the space-time has a naked 
singularity. Since
causality can be violated in Kerr space-times with no event 
horizon~\cite{carter2, chandra}, the assumption of the Cosmic Censorship 
Conjecture appears well motivated.

Interestingly, several solutions of the Einstein equations that lead to the 
formation of naked singularities are known (see e.g. 
Refs.~\cite{chr84, piran, joshi93, chr94, joshi94, joshi98}). 
Moreover, the reliability of GR at the singularity is surely 
questionable. GR is a classical theory and it is widely believed that the 
Planck scale, $E_{Pl} \sim 10^{19}$~GeV, is its natural UV cut-off. New 
physics needs to replace the singularity with something else~\cite{nakao}; 
then the space-time in the full theory may present no pathologies. If this 
is the case, the Cosmic Censorship Conjecture may not be necessary and 
super-spinning Kerr objects with no event horizon, or ``super-spinars'', 
may exist in the Universe~\citep{horava}.

A subtle point is the stability of these objects, but the issue
is not so easy to address, especially because we do not know the actual 
space-time emerging from the full theory. The simplest kind of
instability is represented by the process of accretion itself, which may
reduce $|a_*|$ and convert the super-spinar into a BH~\cite{defelice}. 
However, the evolution of the Kerr parameter 
due to accretion is usually negligible in stellar mass objects in 
binary systems and, as discussed in this paper, the accretion onto
super-spinars is strongly suppressed, because of the presence of 
powerful outflows. A different issue is the intrinsic stability of the 
space-time. A ``true'' Kerr naked singularity is {\it probably} 
unstable~\cite{dotti06, dotti08}, but such a conclusion is based on an 
analysis of the whole Kerr space-time with $- \infty < r < + \infty$, 
which includes ``another Universe'', connected to ``our Universe'' 
through the singularity and admitting closed time-like curves. Another 
danger is represented by the so-called ergo-region 
instability~\cite{cardoso1, cardoso2}. The stability of the space-time 
is, however, mainly determined by the boundary conditions, which are
unknown in our case. The motivation to consider super-spinars is that 
high energy corrections to classical GR may replace the singularity with 
something else and that the space-time has no pathologies in the full 
theory. Since we do not know the full theory, we cannot predict the
actual structure of the space-time at very small radii. Here we do not 
further discuss this point: we simply assume that super-spinars are stable or 
quasi-stable objects. We study their astrophysical implications. Let 
us also notice that, even if super-spinars were unstable, they may 
still be a possible intermediate state of the gravitational collapse,
before decaying into spinning BHs after some characteristic time scale.
Their astrophysical implications may still be very intriguing.

It is thus interesting to look for observational signatures capable of 
distinguishing super-spinars from ordinary BHs, with present and future 
astrophysical experiments. Refs.~\cite{bf09, bft09} discussed the 
implications on the apparent shape. There it was found that, even if the 
bound $|a_*| \le 1$ is violated by a small amount, the shadow cast by 
the super-spinar (i.e. how it blocks light coming to us from an object 
behind it) changes significantly from the BH case: the shadow for the 
super-spinar is about an order of magnitude smaller as well as distorted.  
This distinction can be used as an observational signature in the search 
for these objects.  Based on recent observations at mm wavelength of the 
super-massive BH candidate at the center of the Galaxy~\cite{doeleman},
in~\cite{bf09} one of us speculated on the possibility that it might 
violate the Kerr bound.

The X-ray thermal spectrum emitted by an optically thick and geometrically
thin accretion disk around a super-spinar was instead investigated in~\cite{th10}.
Surprisingly, for any BH with Kerr parameter $a_*^{BH}$, there is a super-spinar
with Kerr parameter $a_*^{SS}$ in the range $[5/3 \, ; \, 8\sqrt{6}/3]$ whose
spectrum is very similar (and practically indistinguishable) from the one of 
the BH. The result is that the X-ray thermal spectrum observed from 
current BH candidates cannot be used to confirm the Kerr bound $|a_*| \le 1$.
However, the contrary is {\it not} true: the X-ray thermal spectrum of a super-spinar
with $1 < |a_*| < 5/3$ can be significantly different from the one of BH.
The disk temperature is higher and the radiation flux larger, even by a few
orders of magnitude for $|a_*|$ slightly larger than 1. In 
principle, this could be used to distinguish super-spinars from BHs.

In this paper, we extend the study on the accretion process onto
super-spinars started in~\cite{bfhty09}. Here we consider the counterpart 
of the Bondi accretion onto BHs. In the case of BHs, this is a quite 
inefficient mechanism to convert the gravitational energy into radiation, 
because a large fraction of the thermal energy is lost behind the horizon. 
The efficiency is
$\eta \sim 10^{-4}$ and thus the Bondi accretion cannot explain many of 
the observed BH candidates. In the case of super-spinars, the efficiency 
can be much higher. As shown in ~\cite{bfhty09}, near super-spinars there
are space regions where the gravitational force is repulsive. In the
Bondi-like case, the gas approaches the compact object with high velocity,
enters the region with repulsive gravitational force, and slows down. 
Some amount of gas, which depends on the spin and the actual size of the
super-spinar, is eventually ejected away, preferably on the equatorial 
plane. Unlike the case of an accretion flow of low temperature and 
low velocity discussed in~\cite{bfhty09}, here the gas reaches the center 
preferably from the axis of symmetry (rather than from the equatorial 
plane) and we find the production of hot outflows, which can have quite
interesting implications. Since outflows produced in the accretion process 
onto BHs are expected to be emitted from their poles, the possible observation 
of powerful equatorial outflows from a BH candidate could be evidence
that such an astrophysical object is actually a super-spinar or, more in general,
a spinning super-compact object with no event horizon.

The paper is organized as follows. In Sec.~\ref{s-pworks}, we briefly summarize 
the results obtained in~\cite{bfhty09}. In Sec.~\ref{s-bondi}, we present 
the results of our simulations of the Bondi-like accretion onto super-spinars.
In Sec.~\ref{s-disc}, we discuss the role of the free parameters of
our model and how their variation changes the accretion process. On the
basis of such numerical study, in Sec.~\ref{s-astro} we discuss a
possible scenario in more realistic cases, mentioning physical and 
astrophysical implications. Summary and conclusions are reported in 
Sec.~\ref{s-concl}. Throughout the paper we use Boyer-Lindquist 
coordinates to describe the Kerr background and natural units 
$G_N = c = k_B = 1$.

\section{Summary of the previous work \label{s-pworks}}

In this section, we briefly review the basic results found 
in~\cite{bfhty09} (see also the discussion in~\cite{b09}). 
There, we studied the accretion process of a test fluid in a background Kerr 
space-time; that is, we neglected the back-reaction of the fluid to 
the geometry of the space-time, as well as the growth in mass and 
the variation in spin of the central object due to accretion. Such an 
approximation can be used if we want to study, for instance, 
the accretion of a stellar 
mass compact object in a binary system, because in this case the matter 
captured from the stellar companion is typically small in comparison 
with the total mass of the compact object. On the other hand, it is not
appropriate to study long-term accretion onto a super-massive object at 
the center of a galaxy, where accretion makes the mass of the compact 
object increase by a few orders of magnitude from its original value.

The calculations were made with the relativistic hydrodynamics module
of the public available code PLUTO~\cite{pluto1, pluto2}, properly 
modified for the case of curved space-time~\cite{banyuls}. 
As equation of state of the accreting matter, we used the one of an 
ideal gas with polytropic index $\Gamma = 5/3$. 
We adopted Boyer-Lindquist
coordinates to describe the Kerr background. The computational domain
was the 2D axisymmetric space $r_{in} < r < 20 \, M$ and $0 < \theta < \pi$, 
where $r_{in}$ was set just outside the event horizon in the case of BH, 
and $r_{in} = 0.5 \, M$ in the case of super-spinar. The choice of 
$r_{in} = 0.5 \, M$ may appear arbitrary, but it was checked that it does 
not significantly alter the final result for any value of $a_*$, as 
long as $r_{in} \lesssim 0.7 \, M$.

We injected gas from the outer boundary into the computational domain
isotropically at a constant rate, with low velocity (specifically,
$v^r = -0.0001$ and $v^\theta = v^\phi = 0$). Because of the simple 
treatment of the accreting matter, the gas temperature was not under 
control. In~\citep{bfhty09}, we simply imposed a maximum temperature: 
the aim was not to find an accurate description of the process, 
but only to capture some characteristic features of the accretion in 
Kerr space-times with $|a_*| > 1$. The code was run with $T_{max} =$ 10~keV, 
100~keV, and 1~MeV, yielding essentially the same result. Such a range 
of $T_{max}$ is the one suggested by observations of galactic BH 
candidates: the hard X-ray continuum (10 -- 200~keV) is a typical feature 
of all these objects and is often explained with a hot inner disk or a 
hot corona, in which the electron temperature is around 100~keV (see 
e.g. Ref.~\citep{liang}). Let us notice that this is not the Bondi 
accretion. The temperature of the gas (ions) at the horizon (in the case 
of BHs) in the Bondi accretion is about 100~MeV.

In~\cite{bfhty09}, we found that the accretion process for $|a_*| > 1$ 
is strongly influenced by the existence of regions near the massive object 
in which the gravitational force is repulsive. Such a repulsive force
is not an inertial effect due to the rotation of the gas, but arises from 
the presence of the naked singularity and is probably a quite common feature
in space-times with naked singularities. In particular, for super-spinars 
with $|a_*| < 1.4$, the accretion process is extremely suppressed and only 
a small amount of the accreting gas can reach the center. Most of the gas 
is accumulated around the object, forming a high density cloud that 
continues to grow. No quasi-steady state was found.

\section{Bondi-like accretion onto super-spinars\label{s-bondi}}

The ``standard'' Bondi accretion describes a spherical, steady-state, 
and adiabatic accretion onto an object. Spherical or quasi-spherical 
accretion flows are 
expected when the compact object accretes from the interstellar medium or when 
it belongs to a binary system in which the companion is massive and has a 
strong stellar wind. The basic features of the Bondi accretion onto a 
Schwarzschild BH can be deduced analytically~\cite{michel} (see also Appendix~G 
of Ref.~\cite{st-book}). The proper velocity of the fluid at the horizon as 
measured by a local stationary observer is equal to 1. For $\Gamma = 5/3$, 
the fluid temperature at the horizon is independent of the temperature at 
infinity and is given by
\be
T_H 
\approx \frac{3}{40} \left(\frac{2}{u_H^2}\right)^{1/3} m
\approx 105 \, \left( \frac{m}{940 \; {\rm MeV}} \right) 
\; {\rm MeV} \, ,
\ee
where $u_H$ is the inward radial component of the 4-velocity at the horizon
($u_H \approx 0.782$ for $\Gamma = 5/3$) and $m$ is the mean baryon mass. The 
accretion luminosity is low, because most of the thermal energy is advected 
into the horizon and lost. The efficiency parameter is $\eta \sim 10^{-4}$, 
which is much lower than the usual value 0.1 -- 0.3 estimated from observations.

The counterpart of the Bondi accretion in the case of super-spinar 
presents some peculiar features. The accretion process near the object 
is clearly far from being spherically symmetric,
as the space-time geometry is only axially symmetric. Moreover,
even the adiabatic description does not work at very small radii, 
because some amount of fluid is heated to very high temperatures. 
Despite that, far from the compact object the conditions are the same 
of the usual Bondi accretion.

With a slightly improved version of the code used in~\cite{bfhty09},
we have studied numerically the 
Bondi-like accretion process onto super-spinars. 
The set-up is almost the same as in Ref.~\cite{bfhty09}.
The default size of the computational domain is still the 2D axisymmetric
space $r_{in} < r < r_{out}$ and $0 < \theta < \pi$, with 
$r_{in} = 0.5 \, M$ and $r_{out} = 20 \, M$. The initial and boundary 
conditions are similar, except for the radial velocity with which
the gas is injected from the computational domain. In the Bondi accretion,
the fluid is almost in free fall and we thus take
\be
v^r \approx - \sqrt{\frac{M}{r_{out}}} \approx -0.224 \, ,
\ee
instead of the very small value adopted in the previous work. As 
in~\cite{bfhty09}, the boundary conditions at $r_{in}$ are such that the
boundary behaves as a perfectly absorbing surface: the gas in the
computational domain can flow to the outside, but no gas can enter the
computational domain from the boundary.

The most important ingredient in the new simulations is that now 
the temperature of the gas is everywhere under control, except in a 
small region at the inner boundary where the repulsive gravitational force 
is stronger. Here the gas temperature becomes too high\footnote{The 
reason is probably that PLUTO works with weak force fields, but finds 
difficulties otherwise. For example, even the public version of the 
code shows a similar behavior in the case of accretion onto a point-like mass 
in Newtonian gravity, when one includes the region close to the object 
(where actually Newtonian gravity should not be used). The same problem 
exists in our simulations of the accretion onto an ordinary BH, if the 
inner boundary is close to the horizon. Nevertheless, since the 
gravitational force around a BH is always attractive, in the latter case
the hot gas does not escape: so we do not have under control its 
temperature in such a region, but there are no effects on the accreting 
gas at larger distances.}. In the Bondi accretion 
onto a Schwarzschild BH, for a fluid with $\Gamma = 5/3$, the temperature
scales as $1/r$ because the gravitational energy is efficiently 
converted to thermal energy. The temperature at the horizon is about 
100~MeV. In this paper we imposed $T_{max} = 1$~GeV as our ``canonical'' value.
For example, the energy of a free particle in the Kerr space-time is
\be
E = \left(1 - \frac{2 M r}{r^2 + a^2 \cos^2\theta} \right) u^t
+ \frac{2 a M r \sin^2\theta}{r^2 + a^2\cos^2\theta} u^\phi \, ,
\ee
where $u^t$ and $u^\phi$ are, respectively, the $t$- and $\phi$-components
of the particle 4-velocity. If the gravitational energy were efficiently
converted to thermal energy, on the equatorial plane the gas temperature
would still roughly scale as $1/r$ and would diverge at the singularity.
For $r = 0.5 \, M$, we would find $T \sim 0.5$~GeV. However, around a
super-spinar the gas can be further heated by the collisions between 
the inflowing and outflowing matter. Thus we set $T_{max} = 1$~GeV. 
We also study the dependence of the accretion process on $T_{max}$ 
in Subsec.~\ref{ss-tmax}. We show that the choice of $T_{max}$ 
does not significantly affect the overall behavior of the flow
around super-spinars.

Let us notice that here we use the same description for the gas in the
whole computational domain. At very high temperatures,
the degrees of freedom, as well as the gas
equation of state, change. Nevertheless we treat the accreting matter
as an ideal gas. Namely, we assume that
the correct qualitative picture of the accretion process can be 
described even neglecting such effects. 
Actually, these uncertainties should be considered together
with the inner boundary conditions.
In our simulations, we take $r_{in} = 0.5 \, M$, 
but the actual size of the massive objects could be much smaller. At very
small distances from the center, the density and the temperature of the gas
can become extremely high, and our physics should not be used. In 
Sec.~\ref{s-astro}, we extrapolate from our numerical study a possible
picture that appears physically reasonable.

The Bondi-like accretion onto super-spinars presents some very
interesting features that were absent in our previous study.
In Ref.~\cite{bfhty09}, the velocity of the gas was low
and only a very small fraction of it was able to enter the region
with repulsive gravitational force. Such a region was indeed almost
empty. Now instead the fluid approaches the massive object with a much higher
velocity, and it thus can enter the region where the net gravitational
force is repulsive. Because of the pressure of the fluid, the gas is
compressed
where the repulsive gravitational force is strong. The gas
is then heated and eventually ejected to larger radii. The amount of gas ejected
and its energy do depend on the spin of the compact object,
i.e. $a_*$, and the temperature of the gas ejected (in our case
$T_{max}$), as discussed in the next section.

\begin{figure}
\par
\begin{center}
\includegraphics[height=8cm,angle=0]{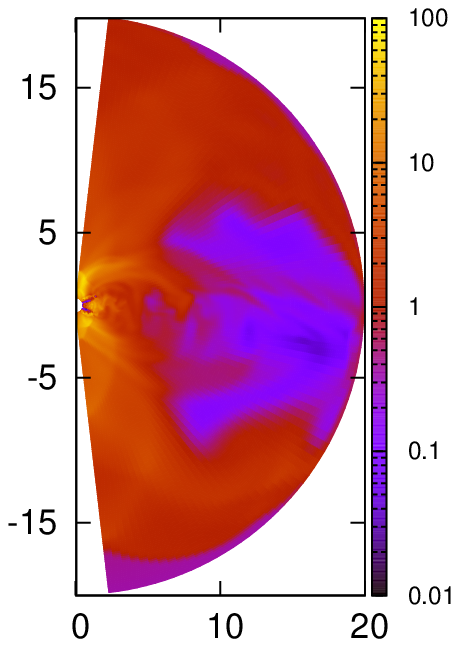} \hspace{.3cm}
\includegraphics[height=8cm,angle=0]{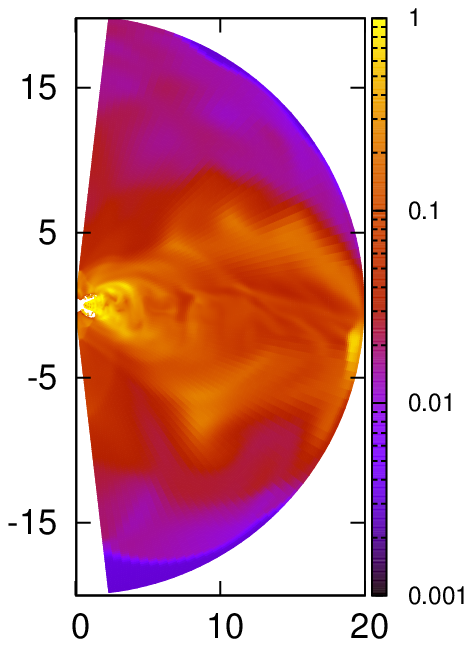} \hspace{.3cm}
\includegraphics[height=8cm,angle=0]{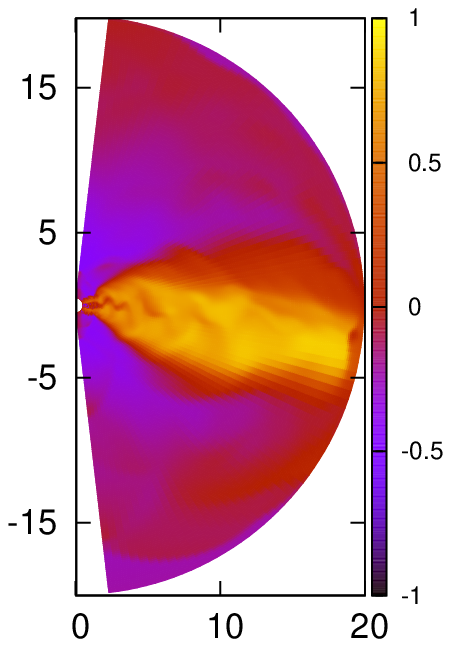} 
\end{center}
\par
\vspace{-5mm} 
\caption{Snapshot at $t = 1000 \, M$ of the density $\rho$ (left 
panels; density in arbitrary units), temperature $T$ (central panel;
temperature in GeV), and contravariant radial component of the 
3-velocity $v^r$ (right panels; velocity in unit $c = 1$) of the 
accreting gas around a super-spinar with $a_* = 1.5$, for 
$T_{max} = 1$~GeV. The unit of length along the $x$ and $y$ axes is 
$M$.}
\label{f-dtv}
\end{figure}

\begin{figure}
\par
\begin{center}
\includegraphics[height=8cm,angle=0]{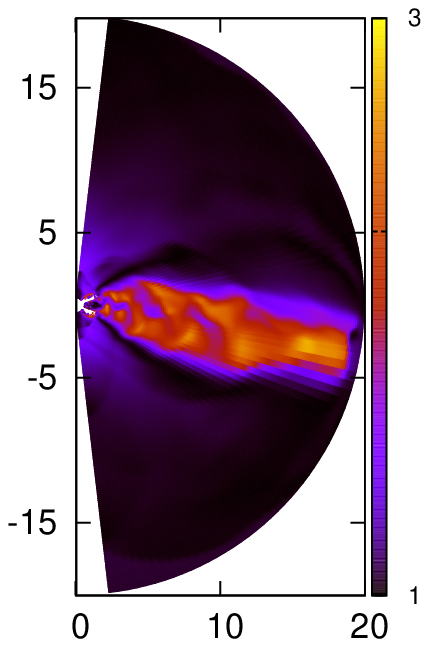} \hspace{.3cm}
\includegraphics[height=8cm,angle=0]{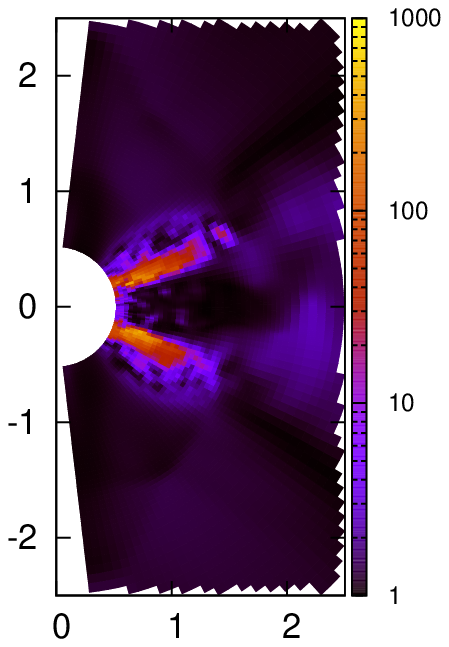} 
\end{center}
\par
\vspace{-5mm} 
\caption{Snapshot at $t = 1000 \, M$ of the Lorentz factor
$\gamma = 1/\sqrt{1 - v^2}$ of the accreting gas around a super-spinar 
with $a_* = 1.5$, for $T_{max} = 1$~GeV. The unit of length 
along the $x$ and $y$ axes is $M$.}
\label{f-gamma}
\end{figure}

\begin{figure}
\par
\begin{center}
\includegraphics[height=8cm,angle=0]{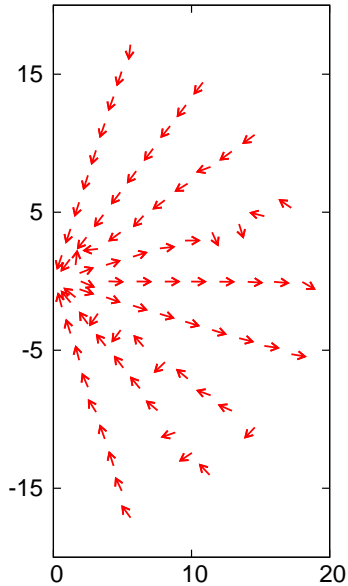} 
\end{center}
\par
\vspace{-5mm} 
\caption{Snapshot at $t = 1000 \, M$ of the direction
of the velocity of the accreting gas around a super-spinar 
with $a_* = 1.5$, for $T_{max} = 1$~GeV. The unit of length 
along the $x$ and $y$ axes is $M$.}
\label{f-arrows}
\end{figure}

\begin{figure}
\par
\begin{center}
\includegraphics[height=7.6cm,angle=0]{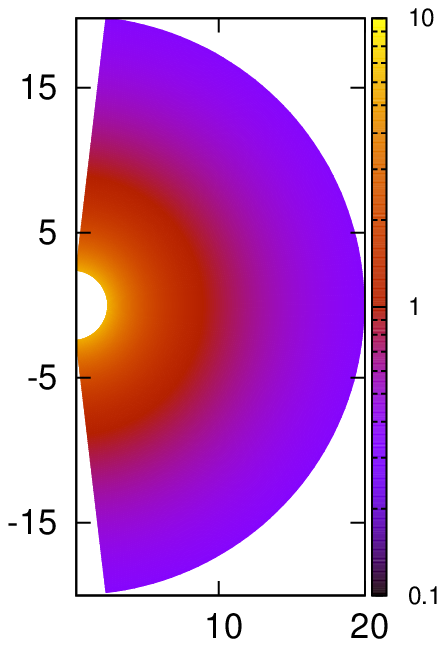} \hspace{.3cm}
\includegraphics[height=7.6cm,angle=0]{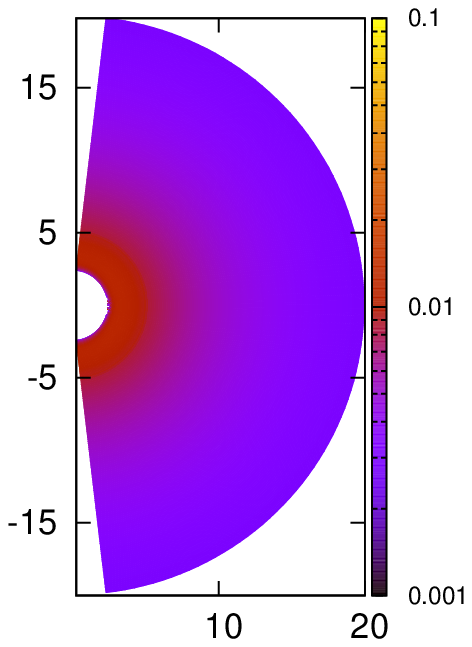} \hspace{.3cm}
\includegraphics[height=7.6cm,angle=0]{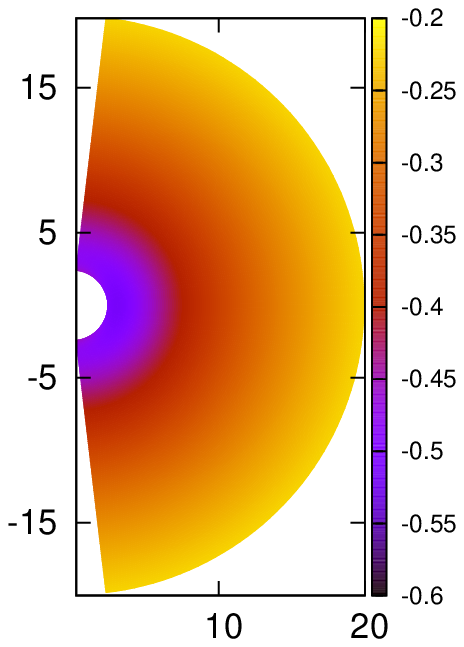} 
\end{center}
\par
\vspace{-5mm} 
\caption{Density $\rho$ (left panels; density in arbitrary units), 
temperature $T$ (central panel; temperature in GeV), and contravariant 
radial component of the 3-velocity $v^r$ (right panels; velocity in 
unit $c = 1$) of the quasi-equilibrium configuration of the accreting 
gas around a Kerr black hole with $a_* = 0.99$. The unit of length 
along the $x$ and $y$ axes is $M$.}
\label{f-dtvbh}
\end{figure}

Density, temperature, and radial velocity of the gas as computed by 
our code are shown in Fig.~\ref{f-dtv} for the case $a_* = 1.5$,
with $T_{max} = 1$~GeV. In Fig.~\ref{f-gamma}, we show the Lorentz
factor $\gamma = 1/\sqrt(1 - v^2)$ and, in Fig.~\ref{f-arrows}, the 
direction of the velocity of the accreting gas in the plane $(r,\theta)$.
Fig.~\ref{f-dtv} can be compared with Fig.~\ref{f-dtvbh},
where we show the same quantities as computed by our code for a Kerr 
BH with $a_* = 0.99$ and $r_{in} = 2.5 \, M$ (no $T_{max}$ is necessary 
here). Since we are discussing the Bondi accretion without the
presence of magnetic fields, in the BH case there are no outflows.

\section{Dependence on $T_{max}$, $r_{in}$, and $a_*$ \label{s-disc}}

\begin{figure}
\par
\begin{center}
\includegraphics[height=5.5cm,angle=0]{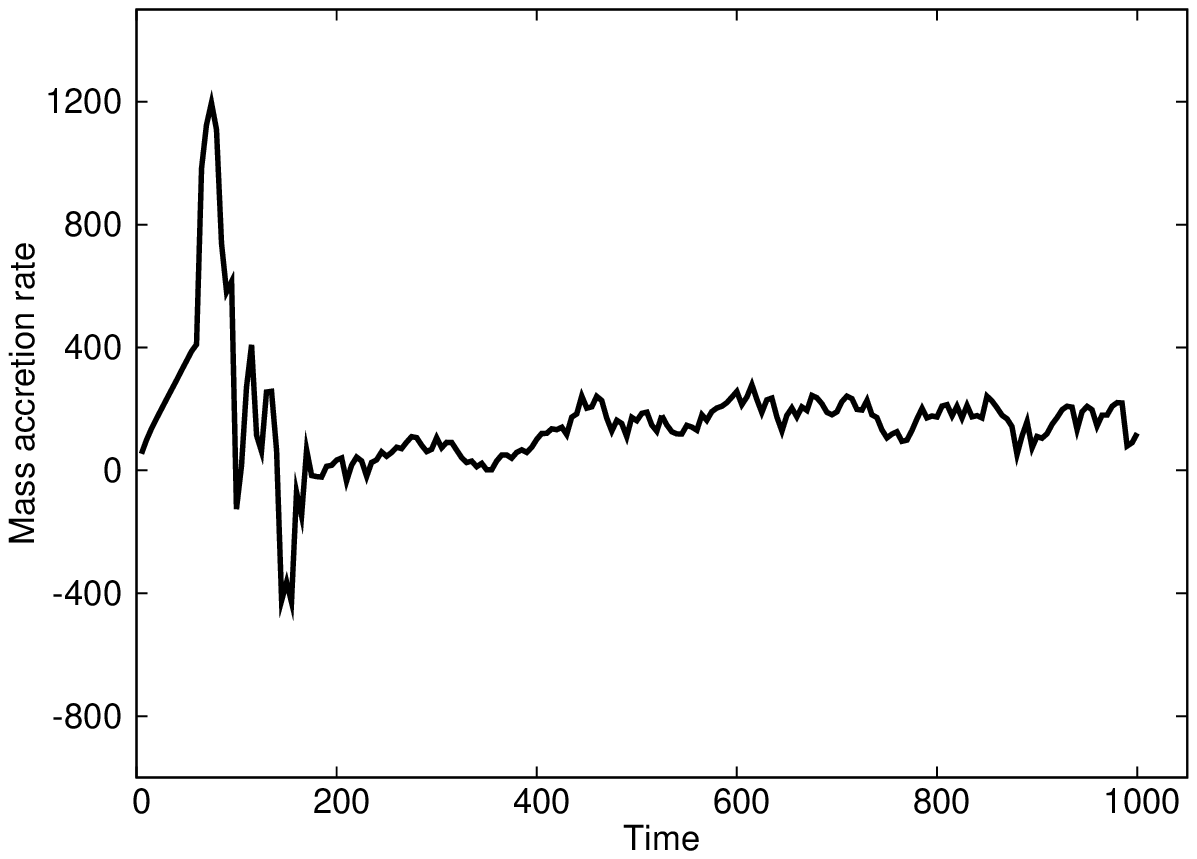} \hspace{.3cm}
\includegraphics[height=5.5cm,angle=0]{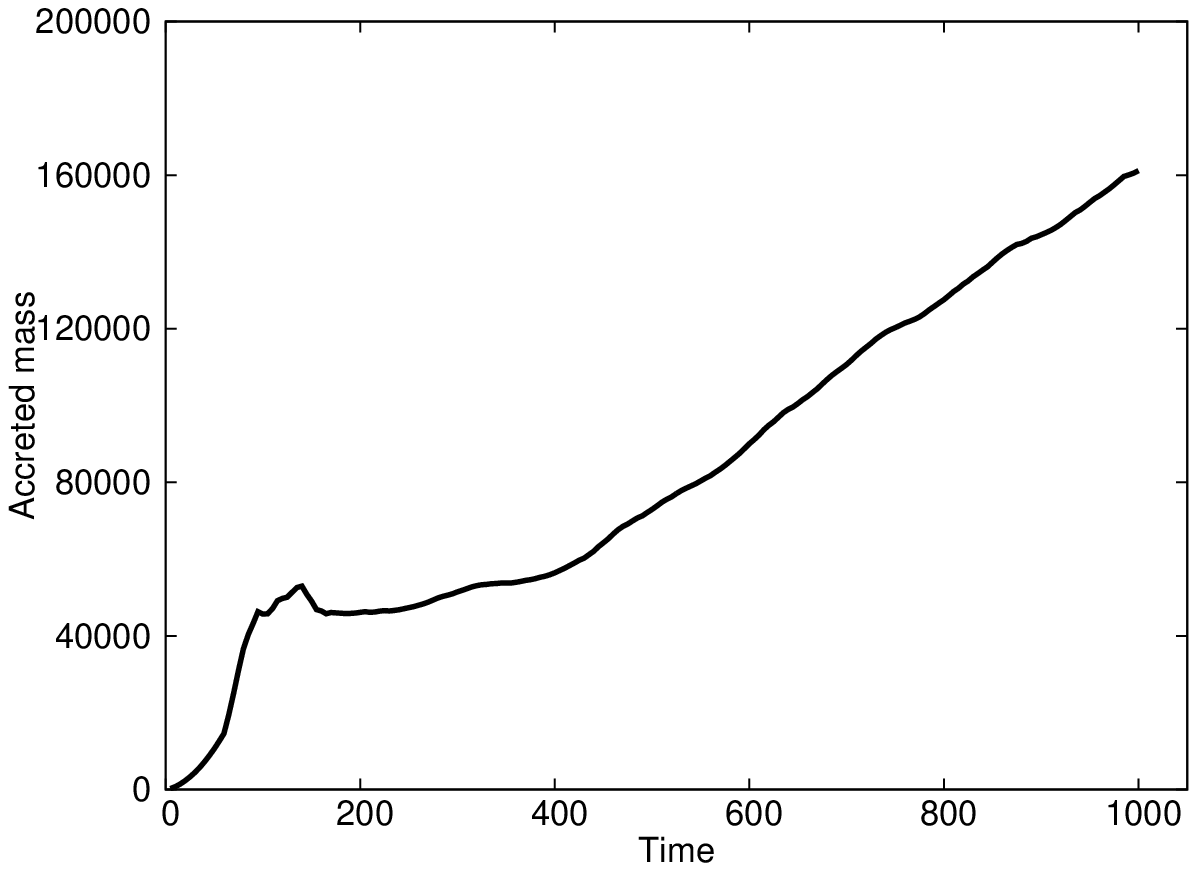} 
\end{center}
\par
\vspace{-5mm} 
\caption{Mass accretion rate (left panel) and accreted mass (right panel)
as a function of time $t$ of the space region inside the radius $r = 5 \, M$
in the case $a_* = 1.5$ and $T_{max} = 1$~GeV. $\dot{M}_{acc}$ and $M_{acc}$
in arbitrary units; $t$ in unit $M = 1$.}
\label{f-default}
\end{figure}

\begin{figure}
\par
\begin{center}
\includegraphics[height=5.5cm,angle=0]{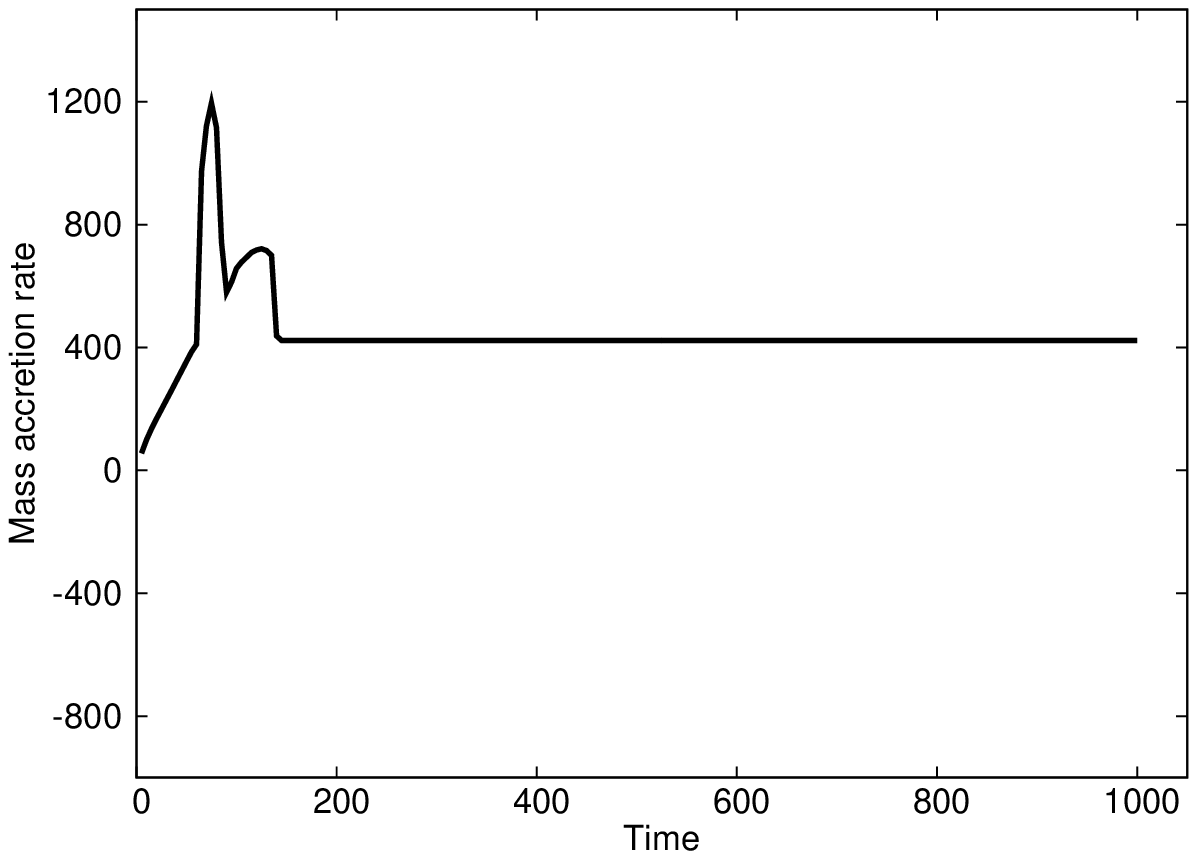} \hspace{.3cm}
\includegraphics[height=5.5cm,angle=0]{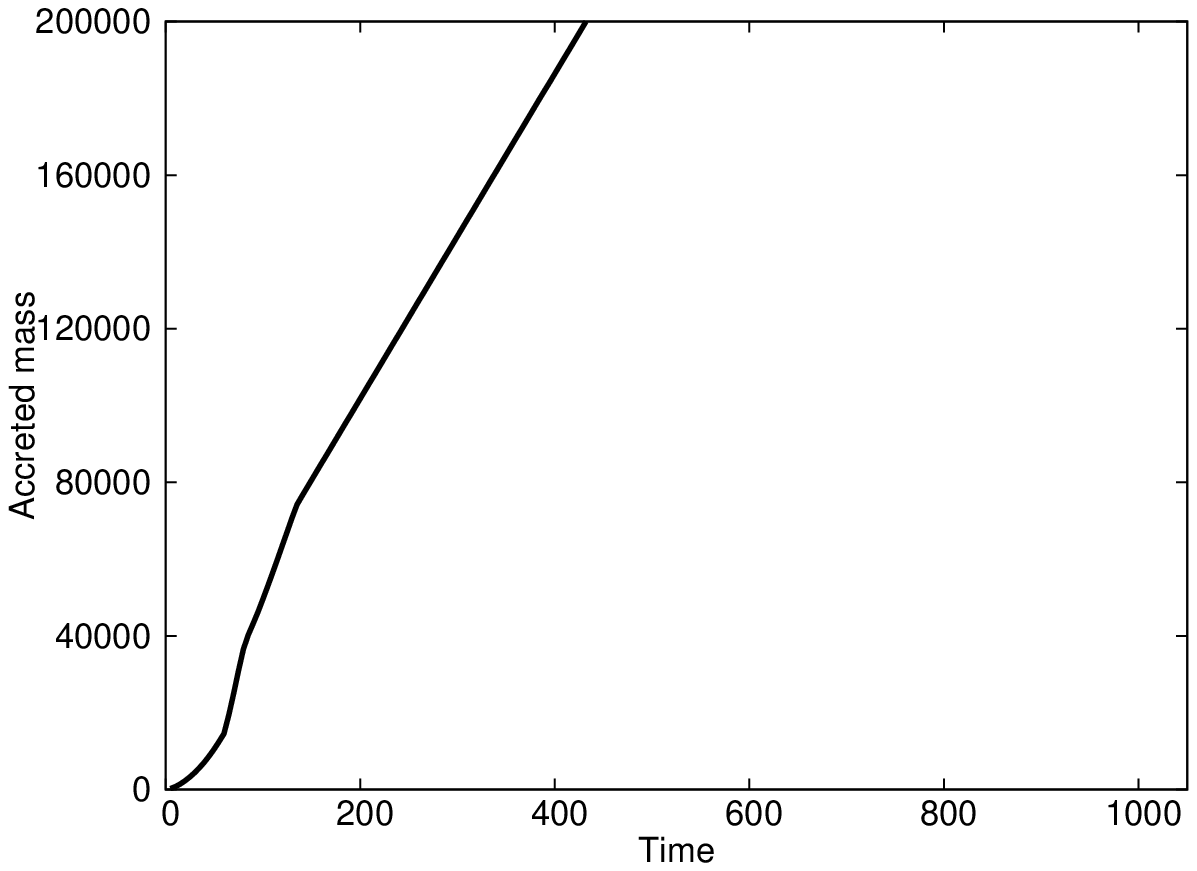} 
\end{center}
\par
\vspace{-5mm} 
\caption{Mass accretion rate (left panel) and accreted mass (right panel)
as a function of time $t$ of the space region inside the radius $r = 5 \, M$
in the case of black hole with $a_* = 0.99$. For $t \gtrsim 150$, the mass 
accretion rate is exactly the amount of mass injected from the boundary.
$\dot{M}_{acc}$ and $M_{acc}$ in arbitrary units; $t$ in unit $M = 1$.}
\label{f-bh}
\end{figure}

The set-up of our simulations is basically characterized by three free
parameters: the maximum temperature $T_{max}$, the inner radius of the 
computational domain $r_{in}$, and the dimensionless Kerr parameter 
$a_*$. In order to elucidate their effects on the accretion process, we 
introduce two physical quantities: the mass accretion rate
of the space inside the radius $r = 5 \, M$, that is,
\be
\dot{M}_{acc}(t) = \int_{r = 5 \, M} \rho v^r 
\sqrt{\gamma} \, d\theta d\phi \, ,
\ee
and its time integral, i.e. the accreted mass of the space inside the 
radius $r = 5 \, M$:
\be
M_{acc}(t) = \int_{t_0}^t \dot{M}_{acc}(\tau) 
\, d\tau \, .
\ee
The temporal behavior of $\dot{M}_{acc}$ and $M_{acc}$ in the interval
$0 < t < 1000 \, M$ are presented in Fig.~\ref{f-default} for our
default choice $a_* = 1.5$ and $T_{max} = 1$~GeV, and can be compared
with the ones for a BH with $a_* = 0.99$, presented in Fig.~\ref{f-bh}.
In the latter case, for $t \gtrsim 150 \, M$, $\dot{M}_{acc}$ is 
exactly the amount of mass injected from the boundary (in our simulations
425, in arbitrary units).

\subsection{Maximum temperature of the gas \label{ss-tmax}}

\begin{figure}
\par
\begin{center}
\includegraphics[height=5.5cm,angle=0]{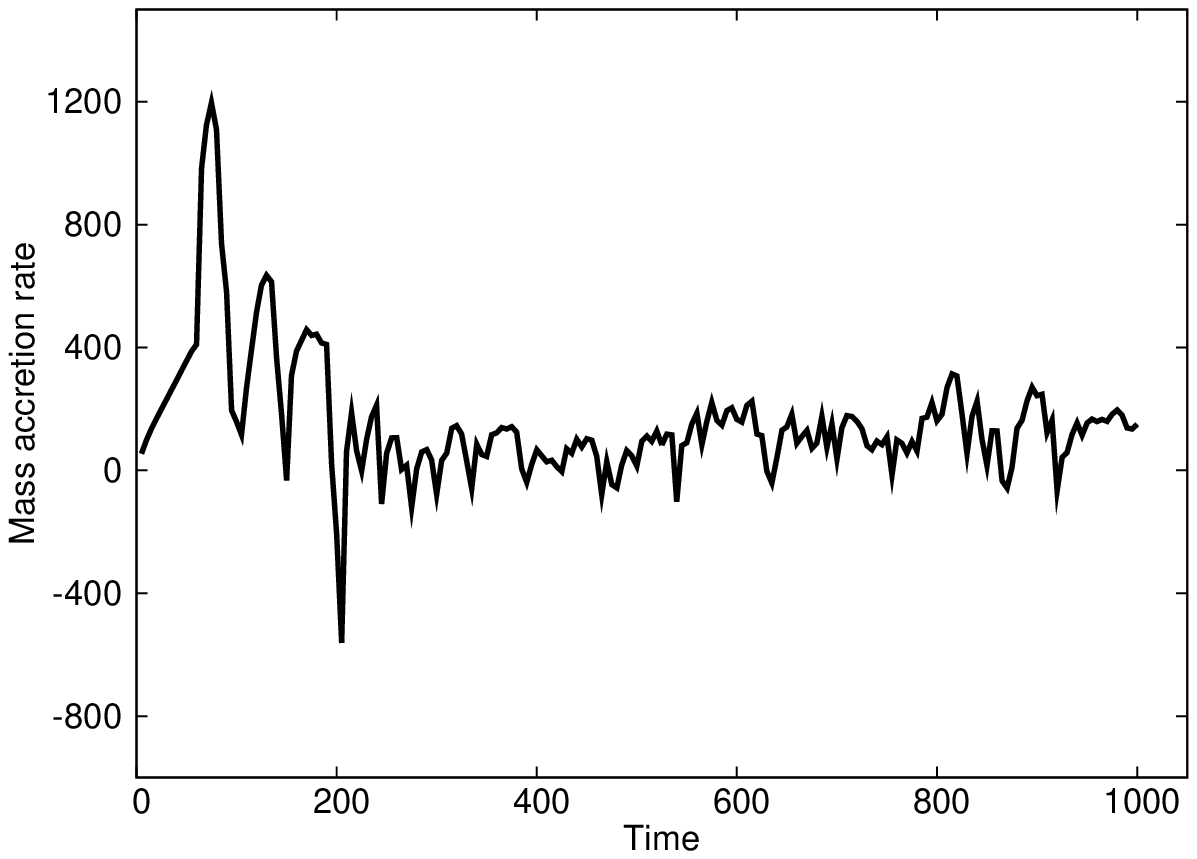} \hspace{.3cm}
\includegraphics[height=5.5cm,angle=0]{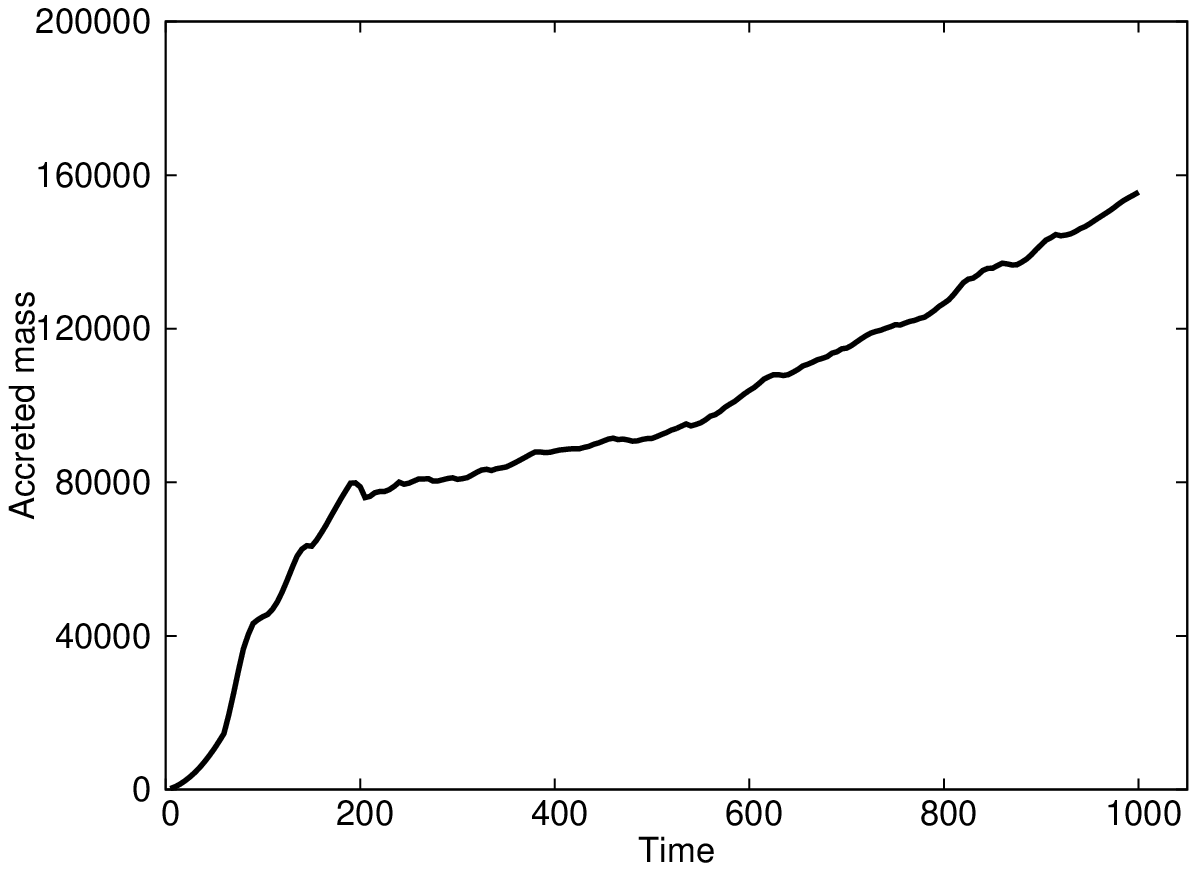} \\ \vspace{.3cm}
\includegraphics[height=5.5cm,angle=0]{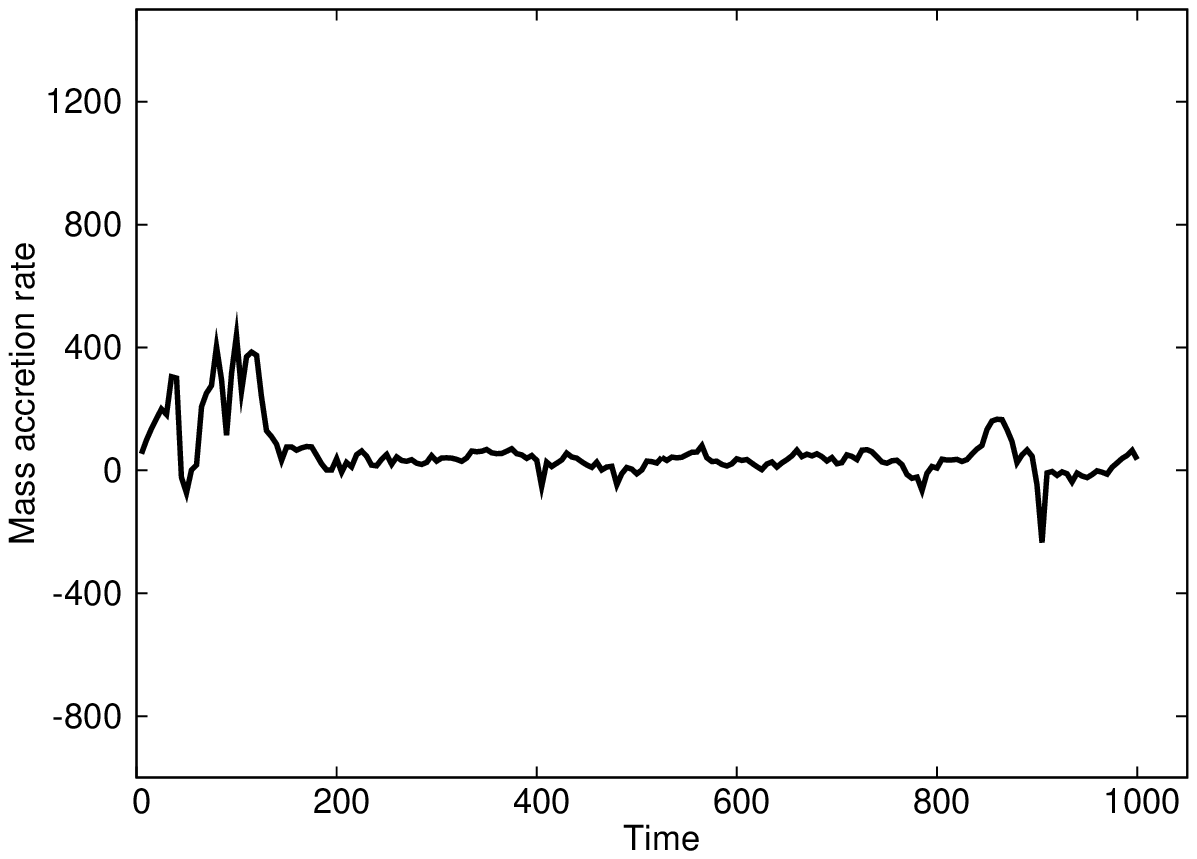} \hspace{.3cm}
\includegraphics[height=5.5cm,angle=0]{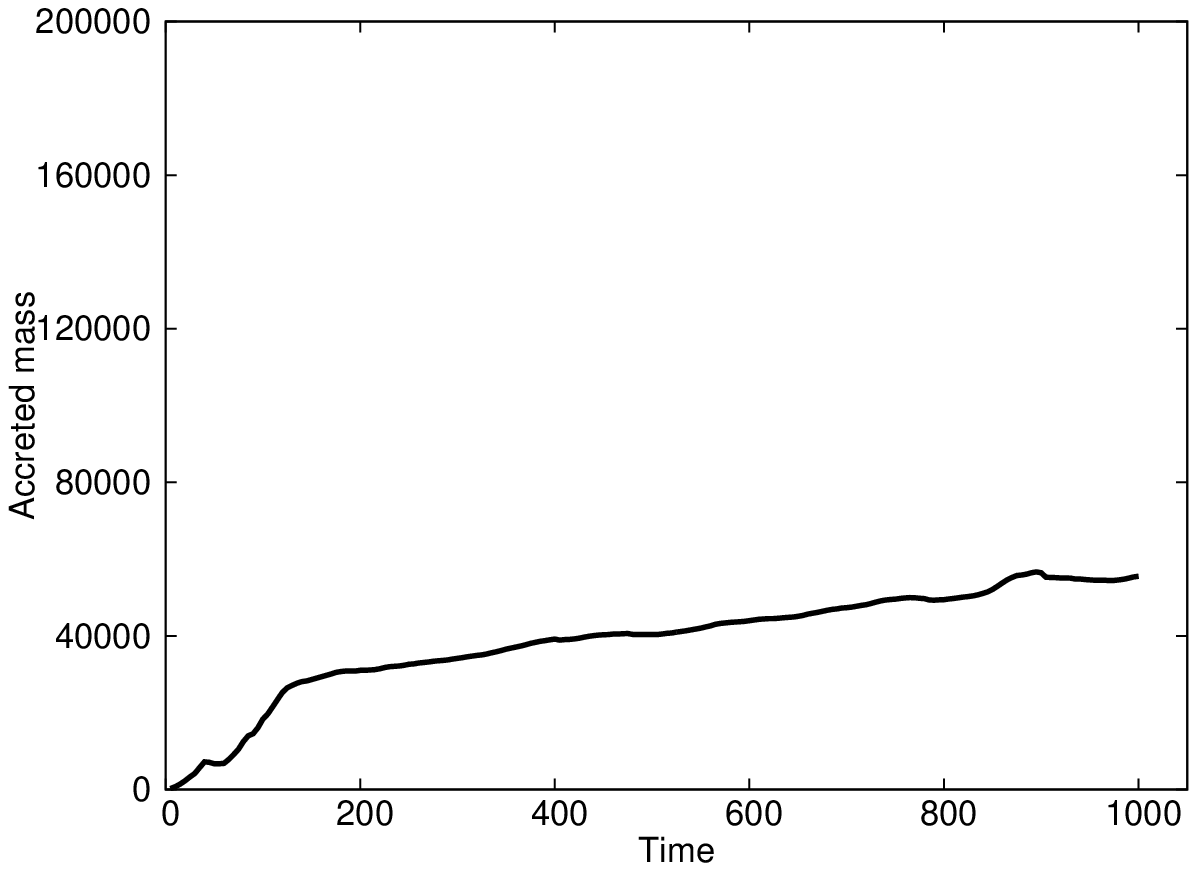}
\end{center}
\par
\vspace{-5mm} 
\caption{Mass accretion rate (left panels) and accreted mass (right panels)
at $r = 5 \, M$ for a Kerr super-spinar with $a_* = 1.5$ for
$T_{max} = 500$~MeV (top panels) and $T_{max} = 2$~GeV (bottom panels). 
$\dot{M}_{acc}$ and $M_{acc}$ in arbitrary units; $t$ in unit $M = 1$.}
\label{f-tmax}
\end{figure}

\begin{figure}
\par
\begin{center}
\includegraphics[height=5.5cm,angle=0]{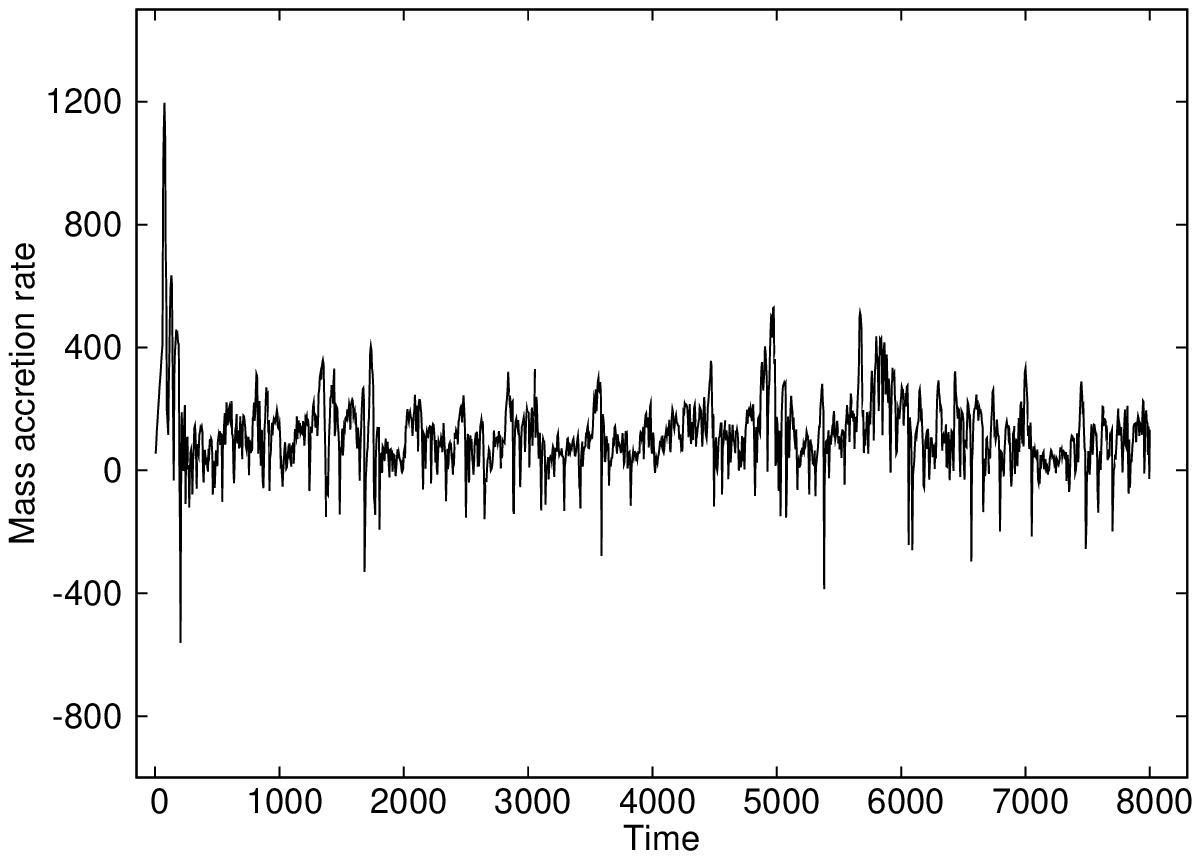} \hspace{.3cm}
\includegraphics[height=5.5cm,angle=0]{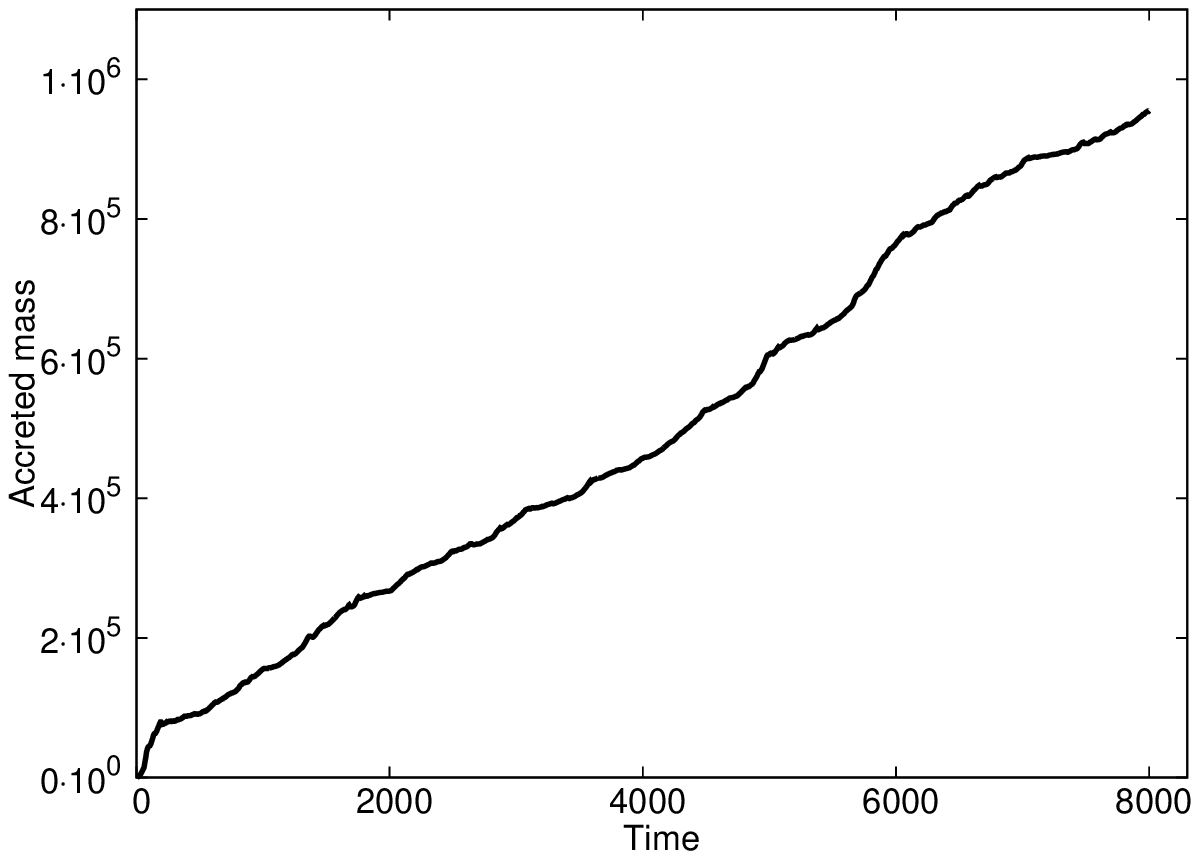} \\ \vspace{.3cm}
\includegraphics[height=5.5cm,angle=0]{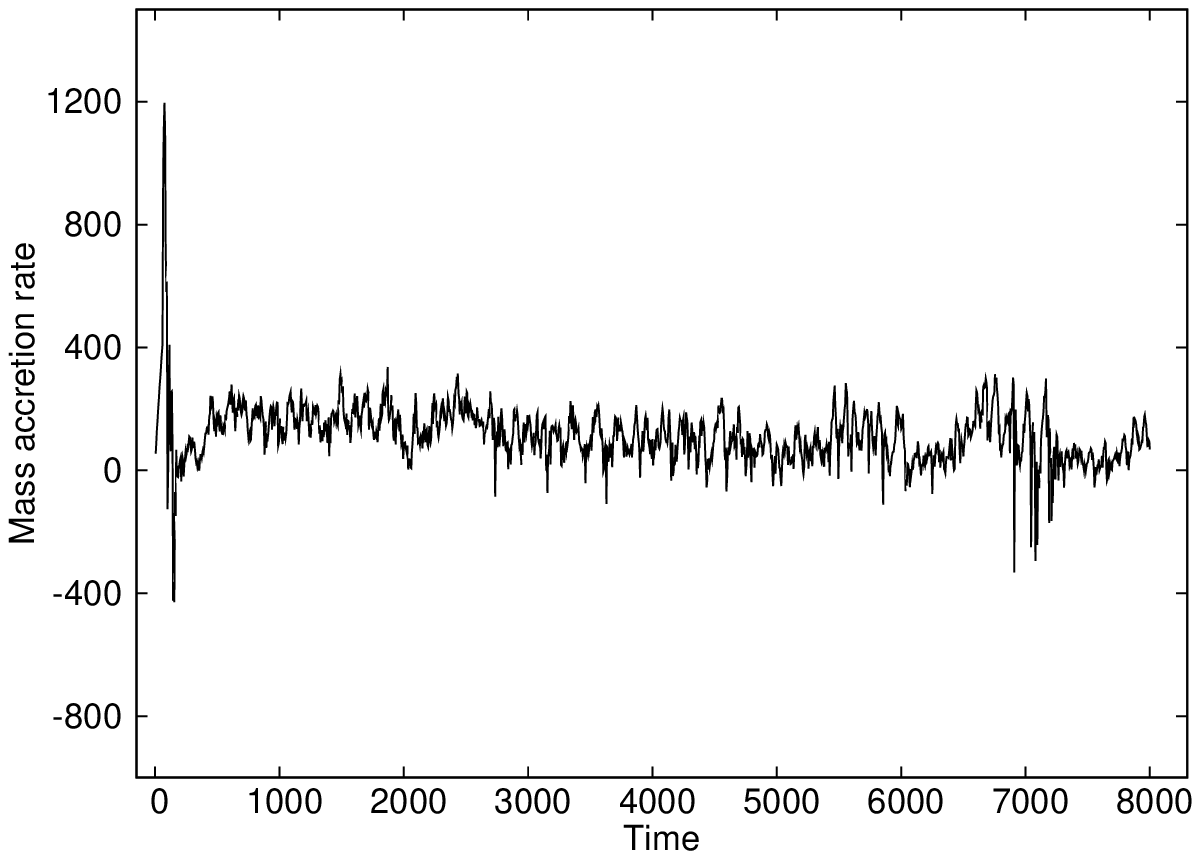} \hspace{.3cm}
\includegraphics[height=5.5cm,angle=0]{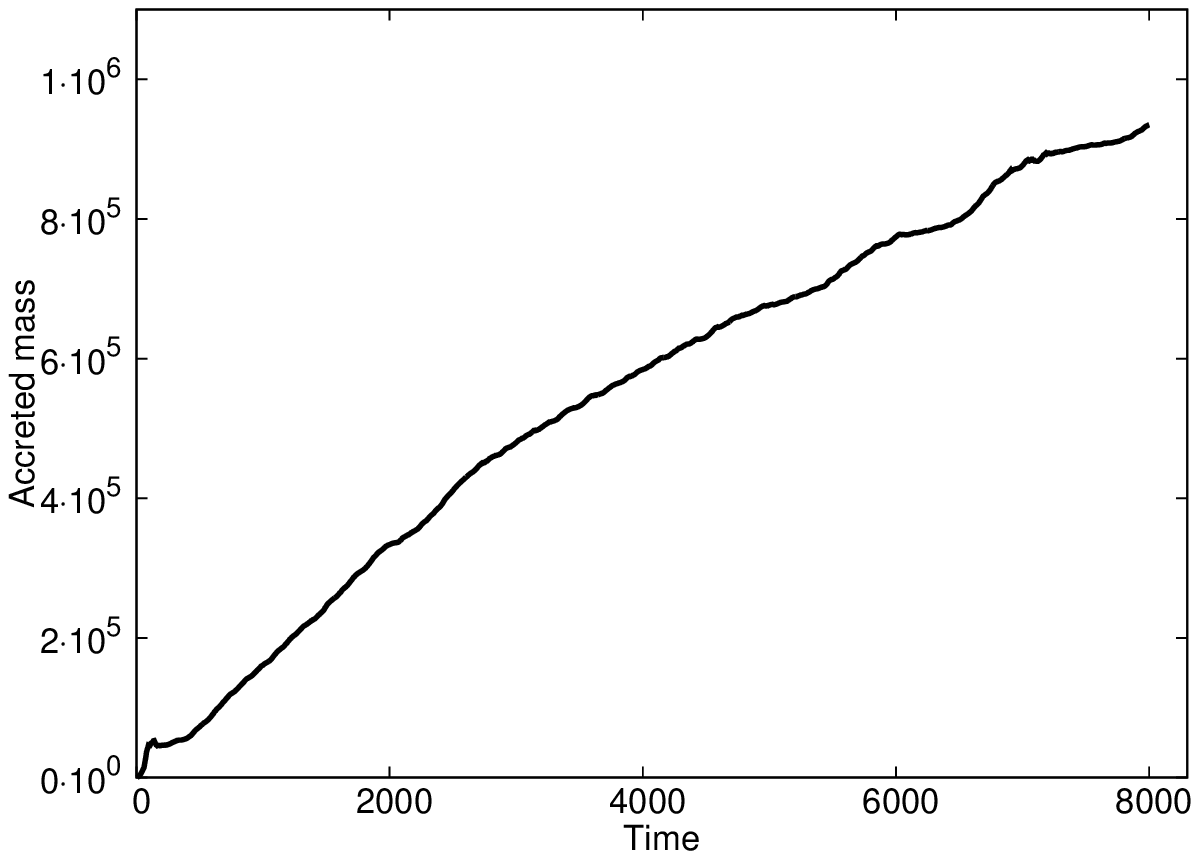} 
\end{center}
\par
\vspace{-5mm} 
\caption{Mass accretion rate (left panels) and accreted mass (right panels)
at $r = 5 \, M$ for a Kerr super-spinar with $a_* = 1.5$ for
$T_{max} = 500$~MeV (top panels) and $T_{max} = 1$~GeV (bottom panels)
up to $t = 8000 \, M$. 
$\dot{M}_{acc}$ and $M_{acc}$ in arbitrary units; $t$ in unit $M = 1$.}
\label{f-tmax-long}
\end{figure}

Our simulations require the imposition of a maximum
temperature in order to avoid unphysical results. 
We thus need to set $T_{max}$. It should depend on physical processes that are 
neglected in our code (e.g. cooling mechanisms), on the choice of 
$r_{in}$, and even on the phenomena occurring at smaller radii (where
actually we cannot predict reliably what happens). It is natural
to expect that a lower (higher) value of $T_{max}$ leads to less
(more) energetic outflows and thus the accretion rate is less (more)
suppressed. This is indeed confirmed by the simulations: in 
Fig.~\ref{f-tmax}, we show the mass accretion rate and the accreted 
mass for $T_{max} = 500$~MeV (top panels) and 2~GeV (bottom panels),
still for $a_* = 1.5$. 
A maximum temperature of 2~GeV at $r = 0.5 \, M$ is likely 
too high, but it is useful to figure out the effect of $T_{max}$ on
the accretion process and check the response of our code.
At the beginning, $0 < t < 200 \, M$, the 
accretion process strongly depends on the initial conditions, but
seems also to give a rough idea of the mass accretion rate at later 
time. As $T_{max}$ increases, the accreted mass at $t = 200 \, M$
decreases. As shown in the next two subsections, the idea that
the accreted mass at $t = 200 \, M$ can suggest how much the accretion
process is suppressed at later time seems to be correct even for 
the other two free parameters, i.e. $r_{in}$ and $a_*$. For $t > 200 \, M$,
the accretion process is more regular and the system reaches 
quasi-equilibrium configurations. In some cases, we observe counterintuitive
results: for example, for $t > 400 \, M$, the mass accretion rate 
for $T_{max} = 500$~MeV is lower than the one with $T_{max} = 1$~GeV (the 
slope of the curve of the accreted mass is less steep). This is 
because the system has not yet reached a quasi-steady state 
at $t = 1000 \, M$. We checked this point by re-running the code for these
two specific cases for a longer time. As shown in Fig.~\ref{f-tmax-long}, 
at later time the accretion process is less suppressed for the case with 
lower temperature, even if the result is not so sensitive to whether we
choose $T_{max} = 1$~GeV or $T_{max} = 500$~MeV. That is good news, 
because it means that the choice of $T_{max}$ is not so important.

\subsection{Radius of the inner boundary}

As the singularity is approached, some observer independent quantities
(e.g. the Kretchmann scalar $K = R_{\kappa\lambda\mu\nu}R^{\kappa\lambda\mu\nu}$) 
diverge. It is usually interpreted as the breakdown of classical GR.
Corrections from new physics become presumably more and more important, and
it is likely that in the full theory there is no central singularity. If deviations 
from the classical Kerr metric are negligible as long as these scalars are 
much smaller than 1 (in Planck units), the simulations should run with 
$r_{in}$ as small as possible. Indeed, according to this criterion, for
astrophysical objects with mass $M$ much larger than the Planck mass,
$M_{Pl} \approx 10^{-5}$~g, the classical metric should be reliable up to 
very small distances from the center, i.e. for $r \ll M$.
However, it is impossible in practice to run the code with $r \ll M$: 
as $r_{in}$ decreases, the computational time increases very quickly. 
In practice, we cannot integrate the system long enough for $r_{in} < 0.3 \, M$. In 
order to catch the effects of a variation of the value of $r_{in}$ on 
the accretion process, we studied what happens if we change the value 
of $r_{in}$. Specifically, we run the code with $r_{in} = 0.3 \, M$, 
$0.8 \, M$, $1.0 \, M$, $1.2 \, M$, $1.5 \, M$, $1.6 \, M$, and $1.7 \, M$,
for $T_{max} = 1$ and 2~GeV.

Roughly speaking, for smaller/larger values of $r_{in}$, the accretion
rate is lower/higher. This is indeed naively expected: at smaller
radii, the density and the pressure of the gas are inevitably higher,
reducing the flow velocity. Moreover, the gravitational force close to
the center is mainly repulsive, which makes it more and more difficult to 
push the gas into smaller and smaller radii.
However, a more detailed study shows the following features\footnote{Here
we describe the case $a_* = 1.5$. Since the spin parameter
determines the region with repulsive gravitational force, for another
choice of $a_*$ one finds a different dependence on $r_{in}.$}. For
$0.3 \, M \lesssim r_{in} \lesssim 1.2 \, M$, there is no significant difference in
the accretion process (adopting the same value for $T_{max}$, while for 
smaller $r_{in}$ one could assume higher maximum temperatures). 
For $1.2 \, M \lesssim r_{in} \le 1.6 \, M$, a slightly higher value of 
$r_{in}$ makes the accretion rate much faster and, for $r_{in} \ge 1.7 \, M$,
the space region with repulsive gravitational force is outside the
computational domain and the accretion process is like the one of BH.

\subsection{Kerr parameter}

In Ref.~\cite{bfhty09}, we discussed an accretion process characterized by 
a gas with low velocity and low temperature. There we found that, for 
small $|a_*|$, the accretion is extremely suppressed: the gas cannot reach 
the central massive object, but is instead accumulated around it, forming a 
high density cloud that continues to grow. As $|a_*|$ increases, a larger
amount of gas can fall to the center and, for $|a_*| \ge 1.4$, a quasi-steady
state configuration exists. In the Bondi-like accretion, the picture is very different,
and, for higher $|a_*|$, the mass accretion rate decreases. That can be explained 
by noting that, for higher $|a_*|$, the repulsive gravitational force close
to the center becomes stronger, thus producing more powerful outflows of gas.
In Ref.~\cite{bfhty09}, the gas was essentially unable to go to the region
where gravity is repulsive, and so we observed neither outflows nor this
dependence on the Kerr parameter.

\begin{figure}
\par
\begin{center}
\includegraphics[height=5.5cm,angle=0]{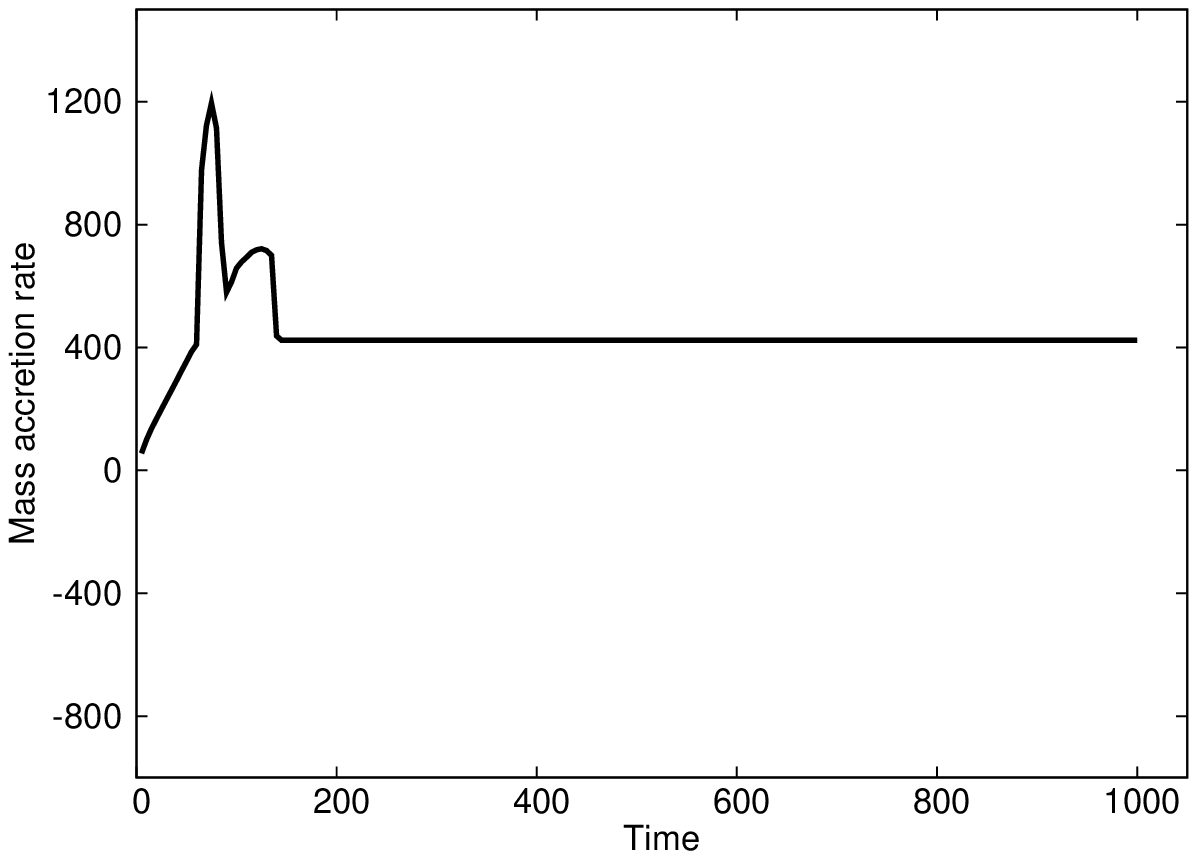} \hspace{.3cm}
\includegraphics[height=5.5cm,angle=0]{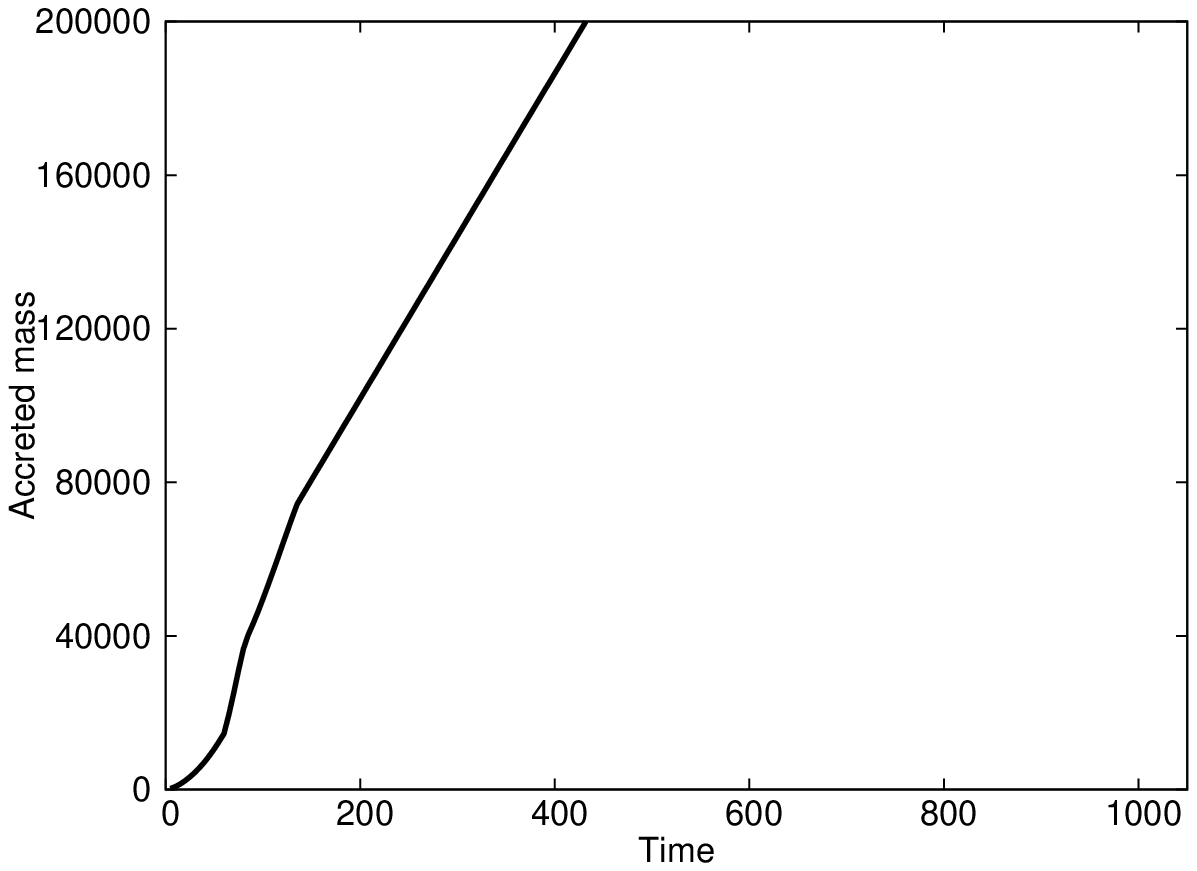} \\ \vspace{.3cm}
\includegraphics[height=5.5cm,angle=0]{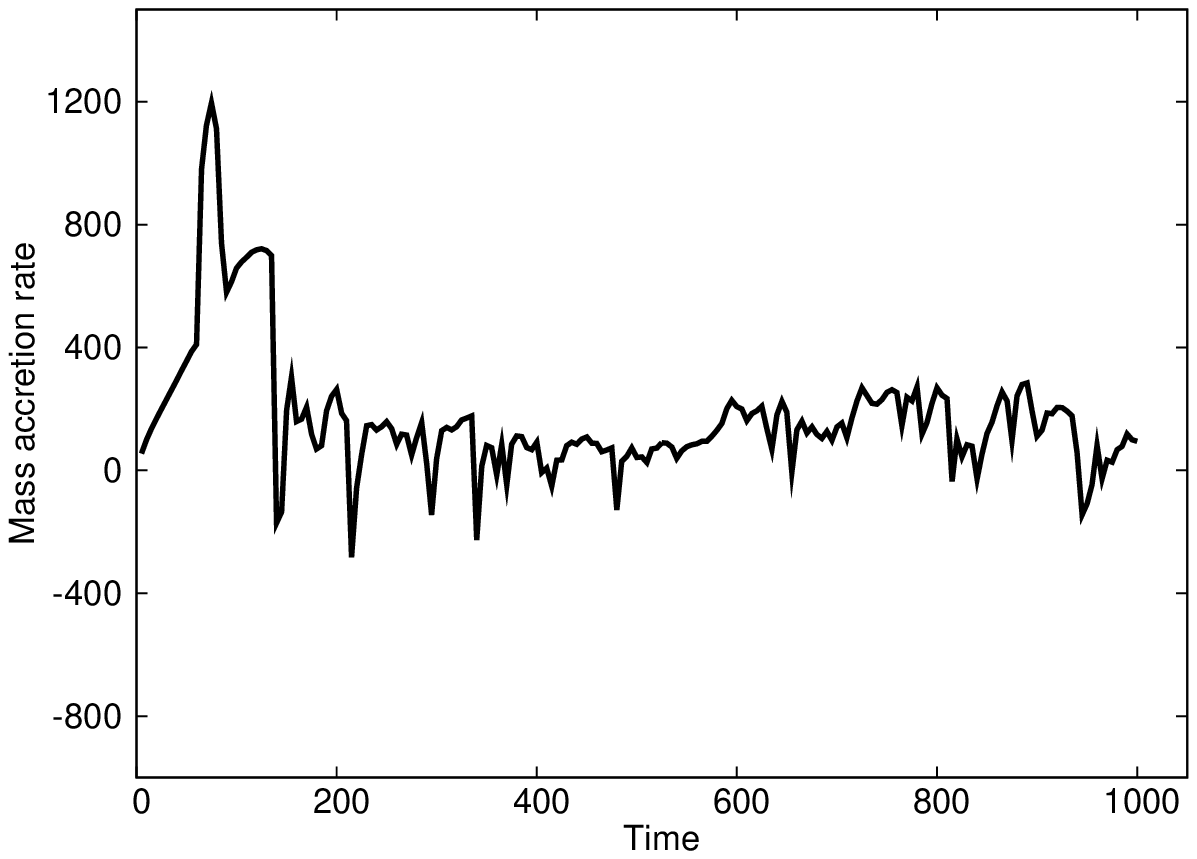} \hspace{.3cm}
\includegraphics[height=5.5cm,angle=0]{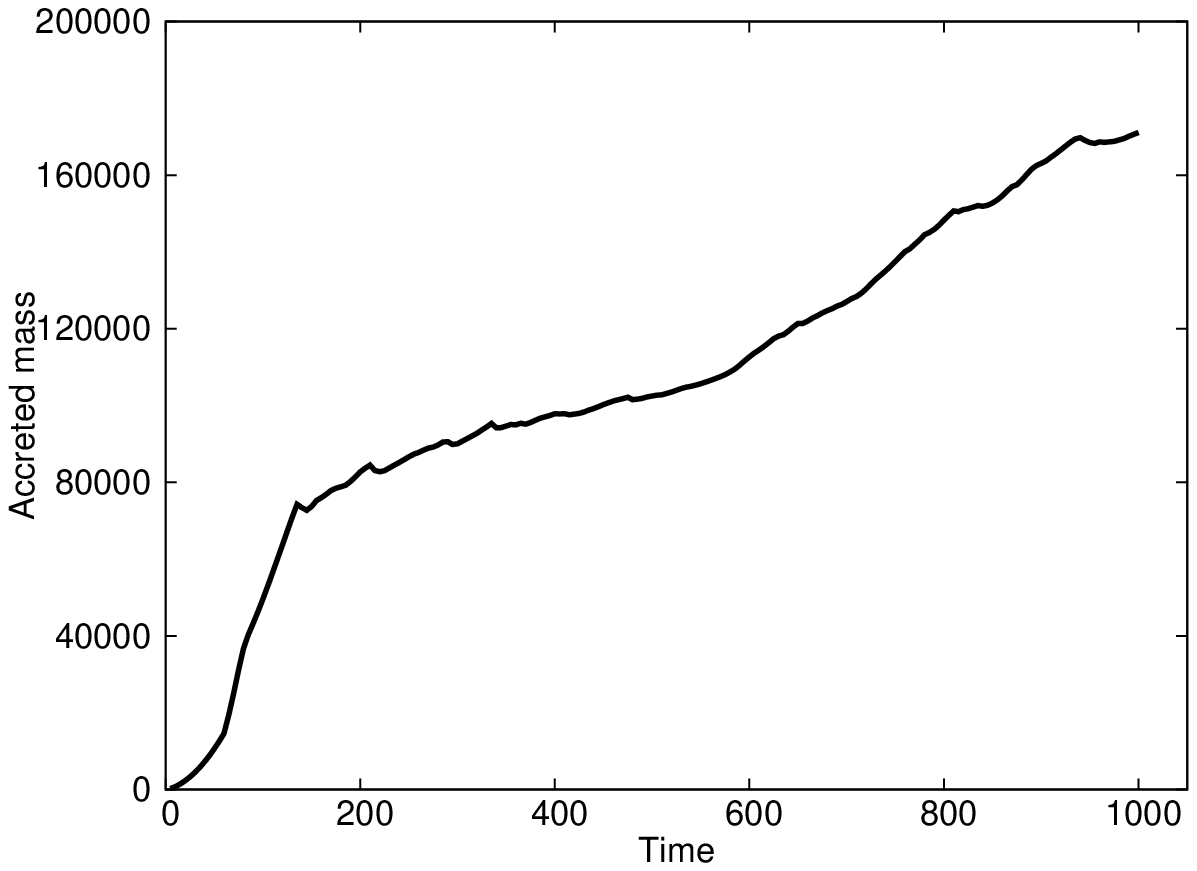} \\ \vspace{.3cm}
\includegraphics[height=5.5cm,angle=0]{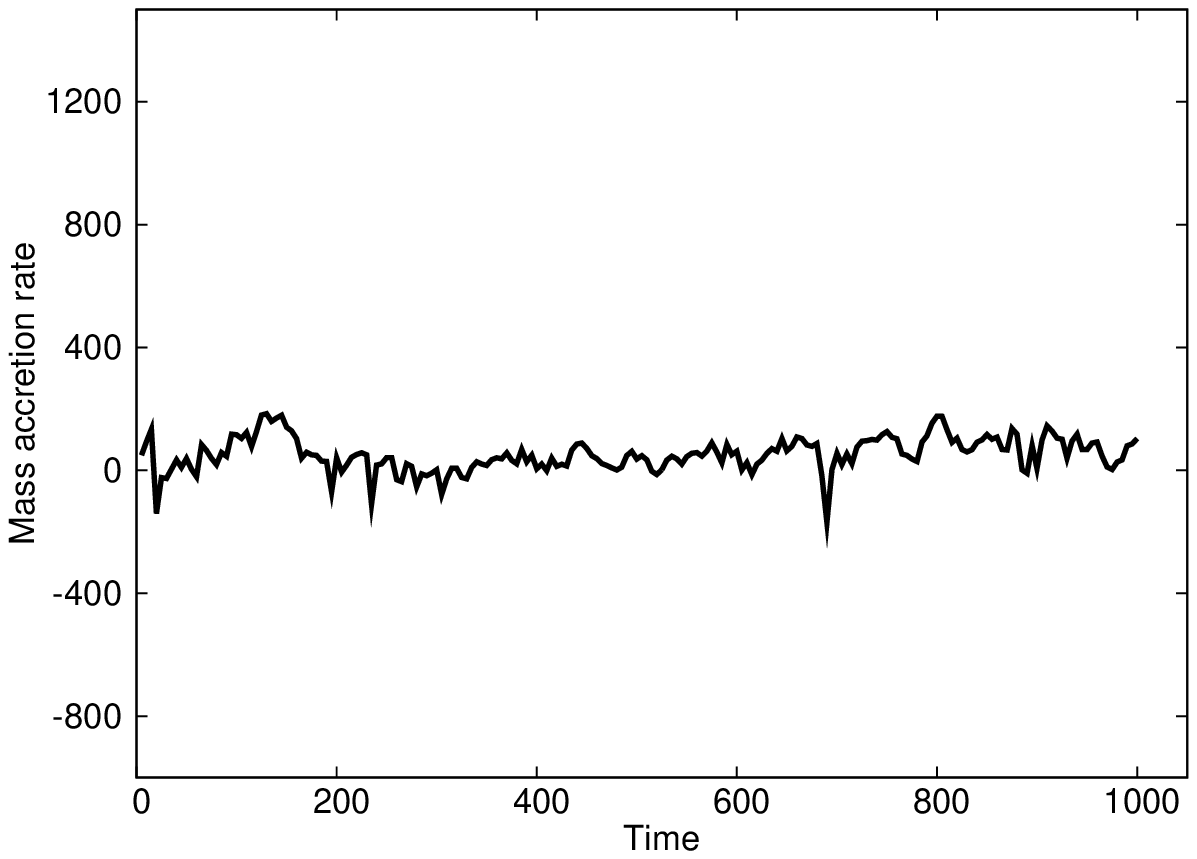} \hspace{.3cm}
\includegraphics[height=5.5cm,angle=0]{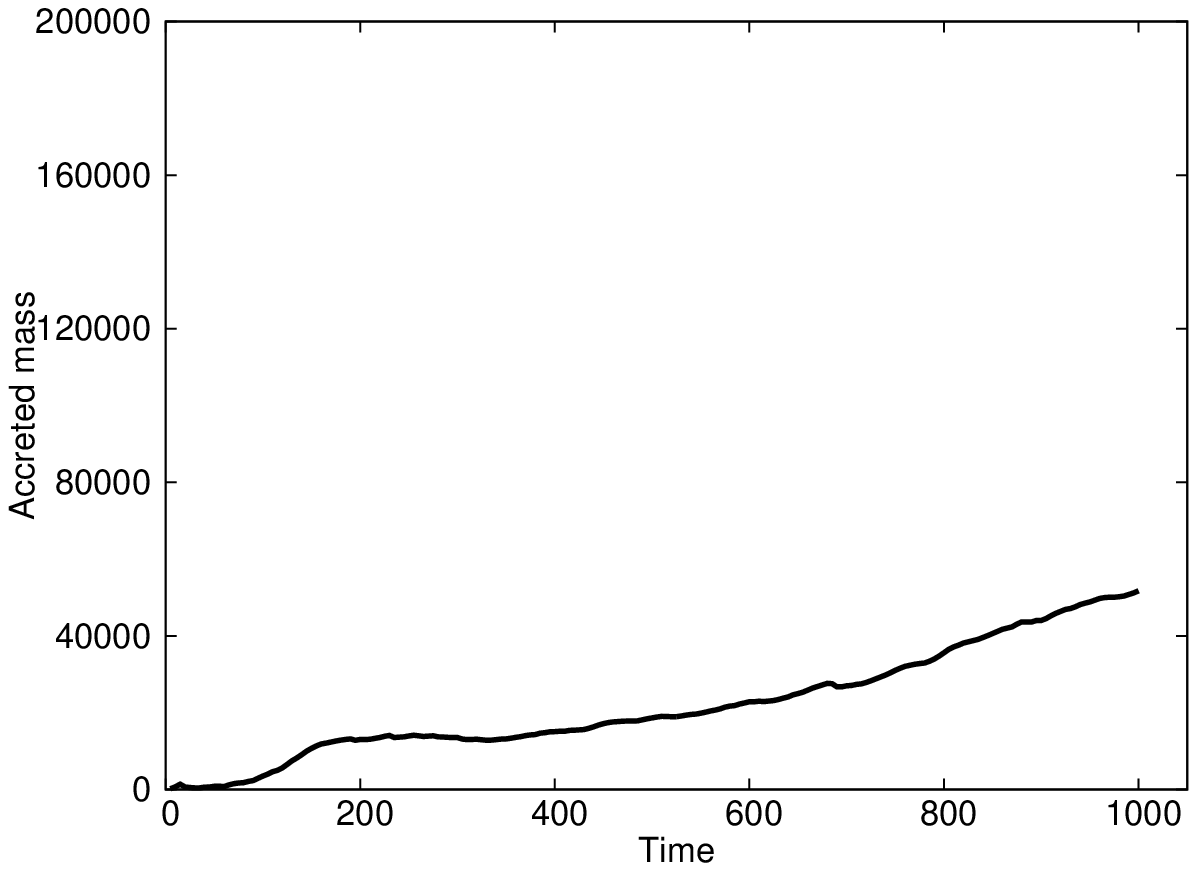} 
\end{center}
\par
\vspace{-5mm} 
\caption{Mass accretion rate (left panels) and accreted mass (right panels)
at $r = 5 \, M$ for a Kerr super-spinar with $a_* = 1.1$ (top panels), 
$a_* = 1.3$ (central panels), and $a_* = 3.0$ (bottom panels).}
\label{f-spin}
\end{figure}

\begin{figure}
\par
\begin{center}
\includegraphics[height=6cm,angle=0]{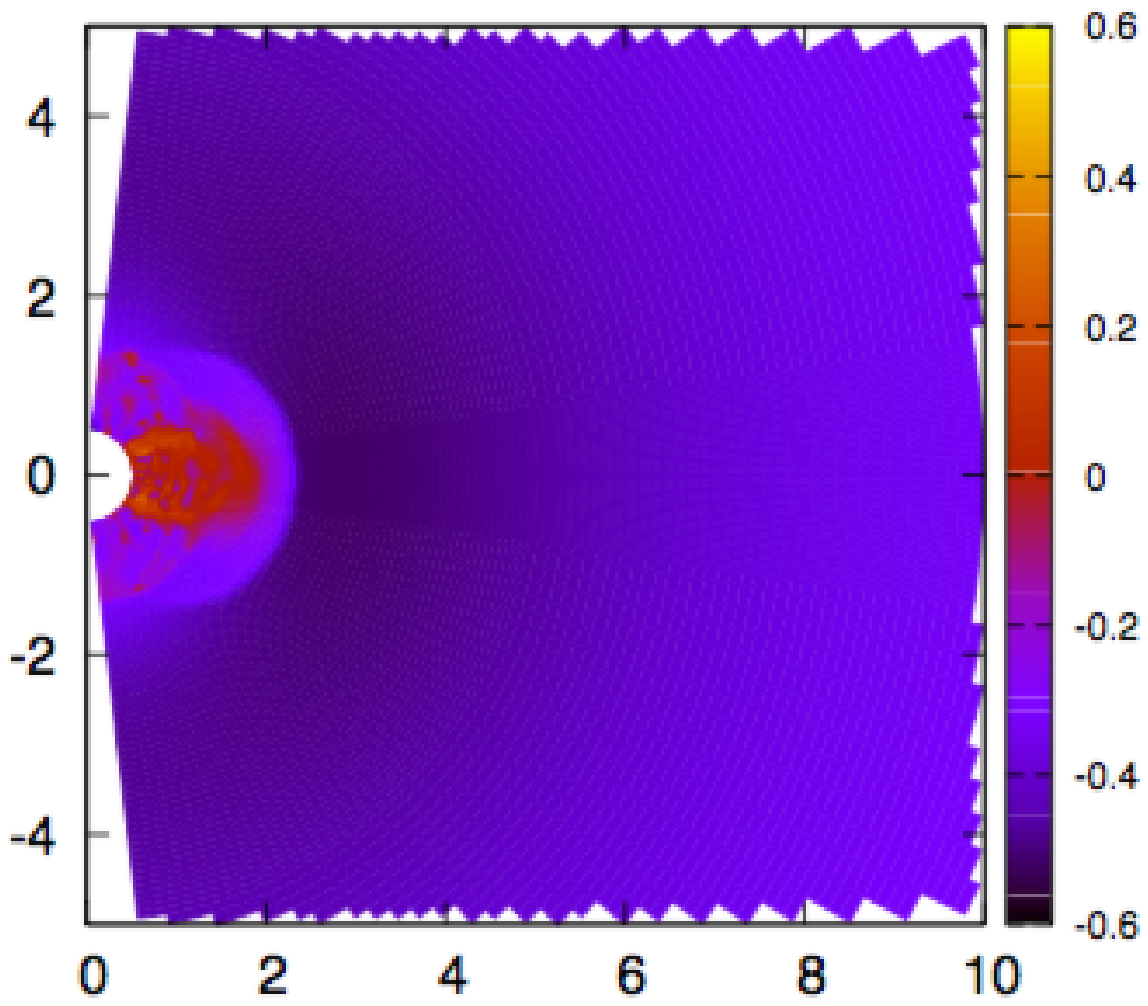} \hspace{.2cm}
\includegraphics[height=6cm,angle=0]{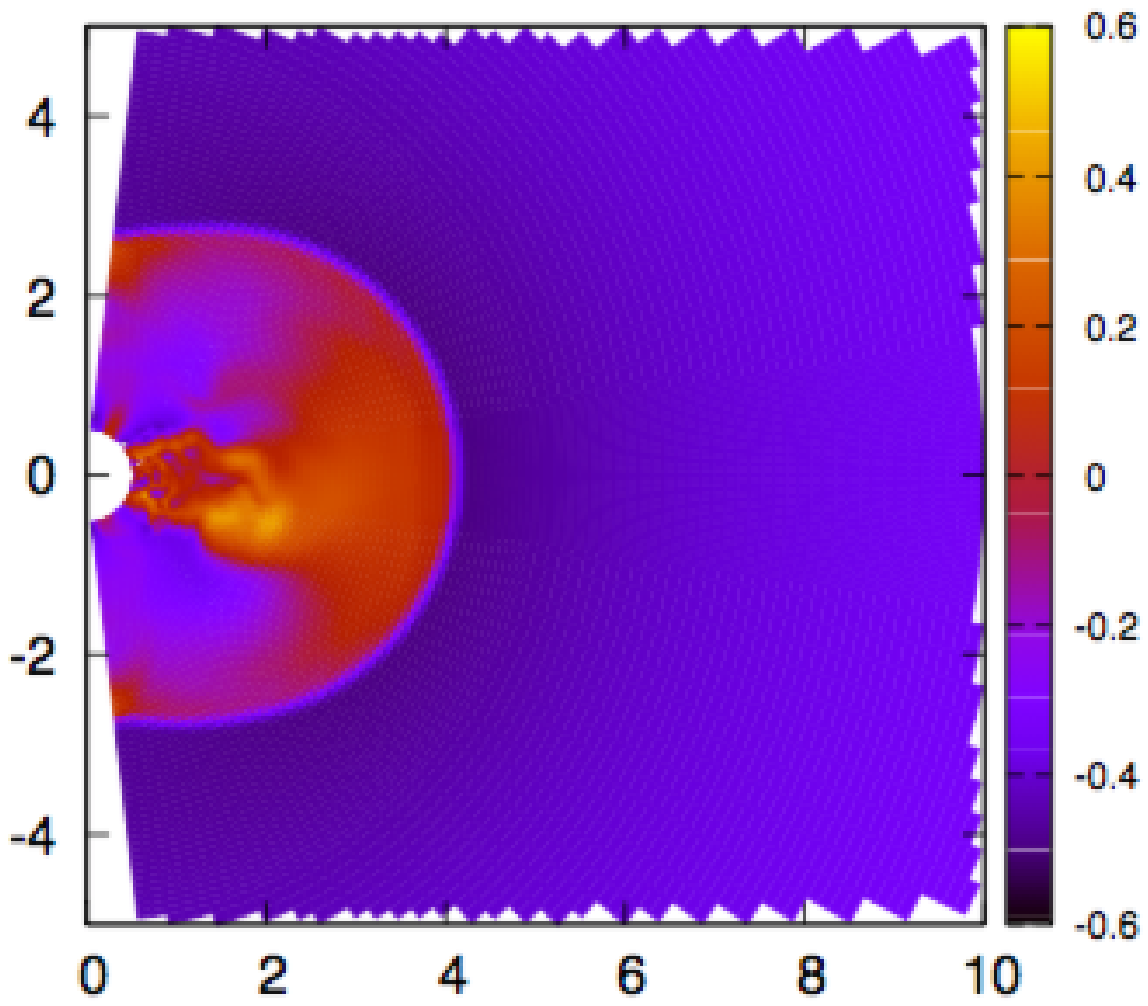} 
\end{center}
\par
\vspace{-5mm} 
\caption{Radial velocity of the accreting gas around a super-spinar with
$a_* = 1.1$ (left panel) and $a_* = 1.2$ (right panel) for $T_{max} = 1$~GeV.
In the first case, the radius of the convective region is about $M$, in
the second case about $3 \, M$.}
\label{f-convect}
\end{figure}

In Fig.~\ref{f-spin}, we show the mass accretion rate and the accreted mass
assuming $T_{max} = 1$~GeV and $r_{in} = 0.5 \, M$. For $a_* = 1.1$ and 1.2, 
the accretion process at $r = 5 \, M$ is similar to the one onto BH of Fig.~\ref{f-bh}. 
Actually there are continuous puffs of hot gas, but their energy is so small 
that this gas is pushed back by the accreting fluid. Specifically, for 
$a_* = 1.1$ the convective region has a radius $r \approx M$, while for 
$a_* = 1.2$ we find $r \approx 3 \, M$; see Fig.~\ref{f-convect}.
There is no gas that leaves the computational domain from the outer boundary, 
exactly like in the BH case. For $|a_*| \ge 1.3$, the outflows
are strong enough to leave the region around the center and eventually
go out of the computational domain at $r = r_{out}$. We checked that the
critical value of the Kerr parameter separating the two different
cases is essentially independent of $r_{in}$ (for fixed $T_{max}$),
while it is sensitive to the value of $T_{max}$; i.e. for higher maximum
temperature the critical value of the Kerr parameter is a bit smaller.
However, it presumably depends on $r_{out}$ and the actual size of the
accreting cloud around the super-spinar. For an infinite size
accreting cloud, we should probably always find a convective region, whose
radius increases for higher values of $|a_*|$ (and $T_{max}$).

\begin{figure}
\par
\begin{center}
\includegraphics[height=4.8cm,angle=0]{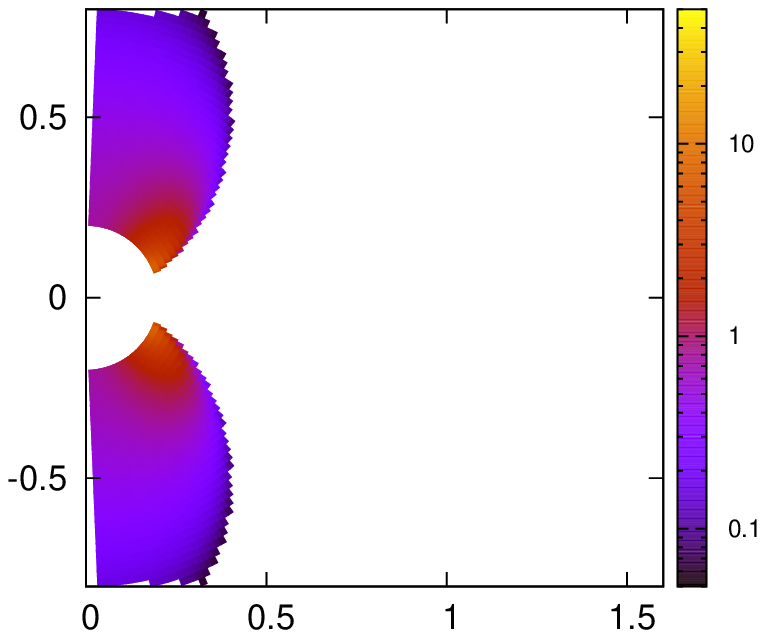} \hspace{.2cm}
\includegraphics[height=4.8cm,angle=0]{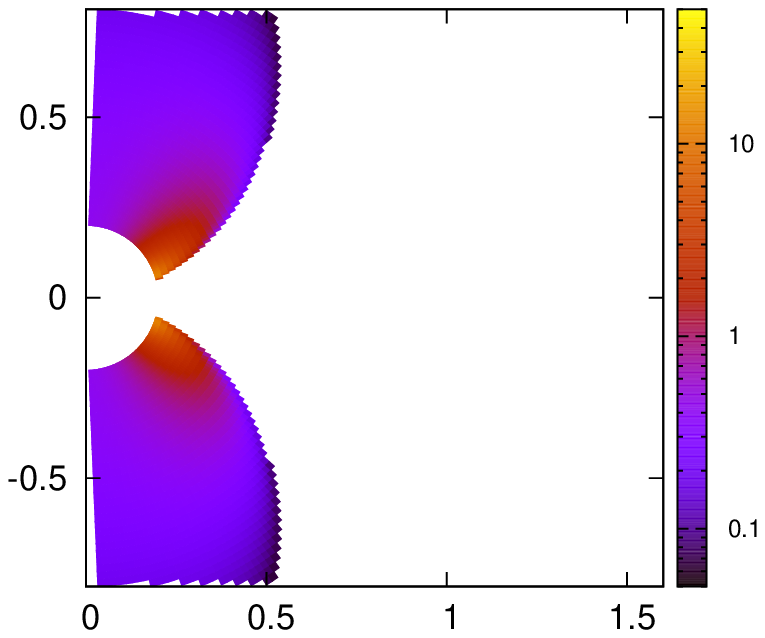} \hspace{.2cm}
\includegraphics[height=4.8cm,angle=0]{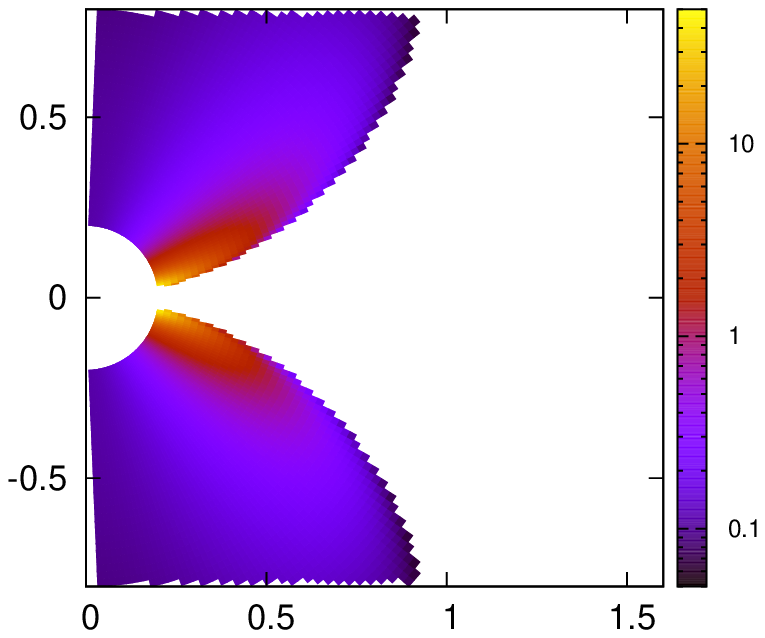} 
\end{center}
\par
\vspace{-5mm} 
\caption{Strength of the repulsive gravitational force for a test-particle
in the region $r > 0.2 \, M$ around a Kerr super-spinar with $a_* = 1.1$
(left panel), $a_* = 1.5$ (central panel), and $a_* = 3.0$ (right panel).
In the white region with $r > 0.2 \, M$, the gravitational force is
attractive.}
\label{f-geod}
\end{figure}

To understand qualitatively the effect of $a_*$ on the accretion process, we
consider the radial component of the geodesic equation of a test-particle.
To get a rough estimate, we can assume $\dot{\theta} = 0$ and thus
\be
\ddot{r} = - \Gamma^r_{tt} \dot{t}^2 
- 2 \Gamma^r_{t\phi} \dot{t} \dot{\phi} - \Gamma^r_{rr} \dot{r}^2
- \Gamma^r_{\phi\phi} \dot{\phi}^2 \, ,
\ee
since all the other terms vanish. $\ddot{r}$ depends on three constants
of motion; that is, the particle energy at infinity $E$, the 
$z$ component of the angular momentum at infinity $L_z$ (where $z$ is the
axis parallel to the spin of the massive object), and the Carter 
constant $\mathcal{Q}$. Let us take $E = m$ (marginally bound orbit;
$m$ is the mass of the test particle) and 
$L_z = \mathcal{Q} = 0$\footnote{In a general Kerr space-time, there is no 
simple interpretation of $\mathcal{Q}$. However, in the Schwarzschild case 
$a_* = 0$, the Carter Constant reduces to $L_x^2 + L_y^2$.}. $\ddot{r}$ in 
the region around the center where the gravitational force is repulsive is 
shown in Fig.~\ref{f-geod}. Intuitively, from these plots one can understand
that higher values of $|a_*|$ produce more energetic and collimated outflows,
and that the gas is preferably ejected around the equatorial plane with
some open angle that depends on $|a_*|$.

\section{Discussion \label{s-astro}}

Let us now discuss the qualitative behavior of the accretion process onto
super-spinars that we can expect in a more realistic situation. To this end,
we need a few assumptions about the space-time at very small radii and the
properties at high temperature of the gas. For example, the naked singularity 
at $r = 0$ might be replaced by a very compact object with a finite radius 
and a surface capable of absorbing the accreting matter\footnote{We warn the 
reader that this assumption may be an important ingredient in our model. It 
is likely one of the simplest and most reasonable options (at least in 
absence of a theory capable of resolving the central singularity), but 
eventually it is responsible for the quasi-steady state configuration with 
strong outflows. If the absorption of gas by the massive object were very 
suppressed, a quasi-steady state configuration would be impossible and 
a cloud would form and grow around the super-spinar.}. Let us also assume 
that around the massive body deviations from the Kerr space-time are not 
significant and that the accreting matter still behaves as a perfect gas. In 
this case, the gravitational force around the object is everywhere repulsive, 
except near the equatorial plane~\cite{bfhty09}. However, as discussed in this
paper, in the Bondi-like accretion onto a super-spinar, the gas cannot
reach the center from the equatorial plane because of the presence of hot
outflows. The result is that the gas approaches the massive object from
the poles. At smaller and smaller radii, the fluid is more and more
compressed, thus reducing further its velocity. Some amount of matter can
presumably reach the surface of the super-spinar and be absorbed, but
most of the matter probably cannot (especially if the size of the 
super-spinar is very small). Since the gravitational force is repulsive, 
the gas cannot pile up indefinitely and some gas is thus ejected to 
larger radii. Outflows of hot low density gas are produced when some amount 
of matter is pushed to the region with strongly repulsive gravitational 
force. Since in the central part of the outflows the gas has a velocity 
exceeding the escape velocity $v_{esc} \sim \sqrt{2M/r}$, see the left panel 
of Fig.~\ref{f-dtv}, some material can go to infinity, at least if the 
finite size of the cloud around the super-spinar is not important in this 
argument. So, we have energetic outflows, which can escape to infinity and 
represent a cooling mechanism for the accreting cloud, and low energy 
outflows, which are ejected to larger radii but are gravitationally 
bound to the super-spinar. If this is really the case, since the cloud of
gas loses energy, eventually the accretion process is not of Bondi type. 
We may thus recover the accretion discussed in Ref.~\cite{bfhty09}:
the formation of hot outflows stops and the gas can fall to the
super-spinar from the equatorial plane, where the gravitational force
is attractive (at least for $|a_*| \ge 1.4$; if $|a_*| < 1.4$ the gas is
accumulated around the super-spinar).

In the Bondi phase, super-spinars are presumably much brighter than
BHs whose efficiency parameter is about $10^{-4}$. Indeed, the
gas is not lost behind an event horizon, but orbits in the
gravitational potential of the massive object. 
There might be the possibility that quite exotic matter is produced 
and ejected to the interstellar medium. Indeed, at very small radii 
the density and the temperature of the gas can be very high and 
inelastic scattering between the particles of the gas may produce 
extremely high-energy, heavy particles. Since blobs of hot gas are 
continuously ejected to larger radii, super-spinars may enrich the 
Universe with these products. If the heavy particles are not highly 
unstable and can decay into Standard Model particles in the 
neighborhood of the super-spinar, the latter might be seen as a 
source of high energy cosmic rays.

Last, we are tempted to put forward the possibility that long
gamma ray bursts (GRBs) might be explained with the formation of
a super-spinar. Of course, one should study the collapse of a
star rather than the accretion process onto a super-spinar, but
our simulations suggest a few features that look quite promising
to produce GRBs. The repulsive gravitational force around 
super-spinars seems to be able to create collimated jets with high
Lorentz factor $\gamma$, as shown in Fig.~\ref{f-gamma}. If these
jets were able to escape from the collapsing envelope, super-spinars
could work as central engines of GRBs. The time variability 
of the accretion rate found in our simulations might suggest that,
for stellar mass super-spinars, even the millisecond time variability
observed in GRBs could be reproduced.

\section{Conclusions \label{s-concl}}

In this paper, we have investigated the counterpart of the Bondi accretion
in the case of super-spinars. Our numerical simulations show that the repulsive 
gravitational force in the neighborhood of the center can produce powerful 
outflows of gas, suppressing the accretion process. It is important to
notice that the gravitational force becomes repulsive already at radii
$r \sim M$ where, for $M \gg M_{Pl}$, classical GR is presumably reliable.
The absence of the horizon can thus lead to the interesting astrophysical
implications discussed in this work.

In our simulations, the
accretion depends on three parameters: the maximum temperature of the gas,
$T_{max}$, the radius of the inner boundary, $r_{in}$, and the dimensionless
Kerr parameter, $a_*$. We have studied numerically how the variation of 
$T_{max}$, $r_{in}$, and $a_*$ changes our results. Then, we have tried to
discuss what can happen in more realistic circumstances that cannot be
simulated by the code. Assuming that the super-spinar can be considered as
a compact object with a finite radius and an absorbing surface, and that
even in its neighborhood deviations from the classical Kerr metric are not
significant, we suggest the following accretion scenario. Only a small fraction
of the accreting gas can really reach the surface of the super-spinar and be
absorbed, since at smaller and smaller radii the gas velocity decreases,
due to two effects: the gravitational force is repulsive and the gas
pressure becomes higher and higher. Most of the accreting gas is pushed back
to larger radii. Some blobs of hot gas are ejected away with high energy
and can escape from the gravitational well of the super-spinar. The rest
of the gas is instead cooled and gravitationally bound to the massive
object. In this way, the cloud of the accreting gas surrounding the
super-spinar cools, less and less gas can enter the region with repulsive 
gravitational force, and eventually the formation of hot outflows stops.
We argue that in this case we recover the accretion process discussed
in~\cite{bfhty09}: the absence of hot outflows let the gas reach the center
from the equatorial plane (for $|a_*| \ge 1.4$, since in this case the
gravitational force is everywhere attractive on the equatorial plane) and
then be absorbed by the surface of the super-spinar.

For astrophysical black holes, it is widely believed that jets and
outflows are powered by magnetic fields (or maybe even by radiation for
high mass accretion rates) and are expected to be ejected from the
poles. It is thus likely that the possible observation of powerful
equatorial outflows from a black hole candidate is a strong
indication that the object is not a black hole, but a
super-spinar or, more in general, a spinning super-compact object
with no event horizon. Regions with repulsive gravitational force
with features similar to the ones shown in Fig.~\ref{f-geod} seem
indeed common in space-times describing the gravitational field
of spinning masses and containing naked singularities.

At least during the Bondi-like accretion phase, super-spinars should be 
objects much brighter than usual black holes, because the gas is not quickly 
lost behind the horizon, but orbits around it. In this phase,
there might also be the possibility of producing exotic matter: scattering 
of the particles near the center, where the density and the temperature
can be very high, may produce heavy particles. Another intriguing 
possibility deserving some attention is that super-spinars might work as 
the central engine for long GRBs. Indeed, they seem to be able to produce 
collimated jets with a high Lorentz factor.


\begin{acknowledgments}
We would like to thank Katherine Freese for collaboration on 
the early stage of this work.
C.B. and N.Y. were supported by World Premier International 
Research Center Initiative (WPI Initiative), MEXT, Japan.
T.H. was partly supported by the Grant-in-Aid for Scientific 
Research Fund of the Ministry of Education, Culture, Sports, 
Science and Technology, Japan (Young Scientists (B) 18740144 
and 21740190).
R.T. was supported by the Grant-in-Aid for Scientific Research 
Fund of the Ministry of Education, Culture, Sports, Science 
and Technology, Japan (Young Scientists (B) 18740144). 
\end{acknowledgments}



\begin{thebibliography}{99}

\bi{hawking}
  S.~W.~Hawking and G.~F.~R.~Ellis,
  {\it The Large scale structure of space-time},
  (Cambridge University Press, Cambridge, UK, 1973).

\bi{penrose}
  R.~Penrose,
  Riv.\ Nuovo Cim.\ Numero Speciale\  {\bf 1}, 257 (1969)
  [Gen.\ Rel.\ Grav.\  {\bf 34}, 1141 (2002)].

\bi{carter}
  B.~Carter,
  Phys.\ Rev.\ Lett.\  {\bf 26}, 331 (1971).

\bi{robinson}
  D.~C.~Robinson,
  Phys.\ Rev.\ Lett.\  {\bf 34}, 905 (1975).

\bi{carter2}
  B.~Carter,
  Phys.\ Rev.\  {\bf 174}, 1559 (1968).

\bi{chandra}
  S.~Chandrasekhar,
  {\it The Mathematical Theory of Black Holes}
  (Clarendon Press, Oxford, UK, 1983).

\bibitem{chr84}
  D.~Christodoulou,
  Commun.\ Math.\ Phys.\  {\bf 93}, 171 (1984).

\bibitem{piran}
  A.~Ori and T.~Piran,
  Phys.\ Rev.\  D {\bf 42}, 1068 (1990).

\bibitem{joshi93}
  P.~S.~Joshi and I.~H.~Dwivedi,
  Phys.\ Rev.\  D {\bf 47}, 5357 (1993)
  [arXiv:gr-qc/9303037].

\bibitem{chr94}
  D.~Christodoulou,
  Annals Math.\  {\bf 140}, 607 (1994).

\bibitem{joshi94}
  I.~H.~Dwivedi and P.~S.~Joshi,
  Commun.\ Math.\ Phys.\  {\bf 166}, 117 (1994)
  [arXiv:gr-qc/9405049].

\bibitem{joshi98}
  S.~S.~Deshingkar, I.~H.~Dwivedi and P.~S.~Joshi,
  Phys.\ Rev.\  D {\bf 59}, 044018 (1999)
  [arXiv:gr-qc/9805055].

\bi{nakao}
  T.~Harada and K.~i.~Nakao,
  Phys.\ Rev.\  D {\bf 70}, 041501 (2004)
  [arXiv:gr-qc/0407034].

\bi{horava}
  E.~G.~Gimon and P.~Horava,
  Phys.\ Lett.\  B {\bf 672}, 299 (2009)
  [arXiv:0706.2873 [hep-th]].

\bi{defelice}
  F.~de Felice,
  Nature {\bf 273}, 429 (1978).

\bibitem{dotti06}
  G.~Dotti, R.~Gleiser and J.~Pullin,
  Phys.\ Lett.\  B {\bf 644}, 289 (2007)
  [arXiv:gr-qc/0607052].

\bibitem{dotti08}
  G.~Dotti, R.~J.~Gleiser, I.~F.~Ranea-Sandoval and H.~Vucetich,
  Class.\ Quant.\ Grav.\  {\bf 25}, 245012 (2008)
  [arXiv:0805.4306 [gr-qc]].

\bibitem{cardoso1}
  V.~Cardoso, P.~Pani, M.~Cadoni and M.~Cavaglia,
  Class.\ Quant.\ Grav.\  {\bf 25}, 195010 (2008)
  [arXiv:0808.1615 [gr-qc]].

\bibitem{cardoso2}
  P.~Pani, V.~Cardoso, M.~Cadoni and M.~Cavaglia,
  arXiv:0901.0850 [gr-qc].

\bi{bf09}
  C.~Bambi and K.~Freese,
  Phys.\ Rev.\  D {\bf 79}, 043002 (2009)
  [arXiv:0812.1328 [astro-ph]].

\bi{bft09}
  C.~Bambi, K.~Freese and R.~Takahashi,
  to appear in the 
  {\it Proceedings of 21st Rencontres de Blois: Windows on the Universe}
  [arXiv:0908.3238 [astro-ph.HE]].

\bi{doeleman}
  S.~Doeleman {\it et al.},
  Nature {\bf 455}, 78 (2008)
  [arXiv:0809.2442 [astro-ph]].

\bibitem{th10}
  R.~Takahashi and T.~Harada,
  Class.\ Quant.\ Grav.\  (in press)
  [arXiv:1002.0421 [astro-ph.HE]].

\bi{bfhty09}
  C.~Bambi, K.~Freese, T.~Harada, R.~Takahashi and N.~Yoshida,
  Phys.\ Rev.\  D {\bf 80}, 104023 (2009)
  [arXiv:0910.1634 [gr-qc]].

\bi{b09}
  C.~Bambi,
  in {\it Proceedings of the Nineteenth Workshop on General Relativity and Gravitation},
  edited by M. Saijo et al., pp. 109-112 (2010) [arXiv:0912.4944 [gr-qc]].

\bi{pluto1}
  A.~Mignone, G.~Bodo, S.~Massaglia, T.~Matsakos, O.~Tesileanu and C.~Zanni,
  Astrophys.\ J.\ Suppl.\  {\bf 170}, 228 (2007)
  [arXiv:astro-ph/0701854].

\bi{pluto2}
  \verb|http://plutocode.to.astro.it/index.html|

\bi{banyuls}
  F.~Banyuls, J.~A.~Font, J.~M.~Ibanez, J.~M.~Marti and J.~A.~Miralles,
  Astrophys.\ J.\  {\bf 476}, 221 (1997).

\bibitem{liang}
  E.~P.~Liang,
  Phys.\ Rep.\  {\bf 302}, 67 (1998).

\bibitem{michel}
  F.~C.~Michel,
  Astrophys.\ Space Sci.\  {\bf 15}, 153 (1972).

\bibitem{st-book}
  S.~L.~Shapiro and S.~A.~Teukolsky,
  {\it Black holes, white dwarfs, and neutron stars: The physics of compact objects},
  (Wiley, New York, New York, 1983).

\end{thebibliography}
\end{document}